\documentclass[a4paper, 11pt]{article}
\usepackage{jheppub}

\usepackage{amsmath}
\usepackage{amsfonts}
\usepackage{amssymb}
\usepackage{float}
\usepackage{multirow}

\usepackage[latin1]{inputenc}

\usepackage{comment}

\usepackage{hyperref}

\usepackage{color}
\usepackage{graphicx}

\usepackage{chngcntr}
\counterwithin*{equation}{section}

\renewcommand{\thefootnote}{\fnsymbol{footnote}}
\renewcommand{\thanks}[1]{\footnote{#1}}

\newcommand{\bea}{\begin{align}}
\newcommand{\eea}{\end{align}}
\newcommand{\ee}{\end{equation}}
\newcommand{\be}{\begin{equation}}
\newcommand{\<}{\langle}
\renewcommand{\>}{\rangle}

\def\op{\ensuremath{\mathcal O}}
\def\hr{\hat{r}}
\def\ha{\hat{a}}
\def\hb{\hat{b}}
\def\hc{\hat{c}}
\def\hu{\hat{u}}
\def\hA{\hat{A}}
\def\hB{\hat{B}}

\def\a{\alpha}
\def\b{\beta}
\def\c{\gamma}
\def\d{\delta}
\def\e{\epsilon}           
\def\g{\gamma}

\def\k{\kappa}                    
\def\l{\lambda}
\def\m{\mu}
\def\n{\nu}
\def\o{\omega}  \def\w{\omega}
\def\r{\rho}                                     
\def\t{\tau}

\def\x{\xi}

\def\L{\Lambda}
\def\O{\Omega}

\def\S{\Sigma}

\def\X{\Xi}
\def\del{\partial}              
\let\a=\alpha \let\b=\beta \let\g=\gamma \let\d=\delta \let\e=\epsilon
    \let\k=\kappa
\let\l=\lambda \let\m=\mu \let\n=\nu \let\x=\xi \let\r=\rho
 \let\t=\tau   \let\c=\chi 
\let\w=\omega    \let\L=\Lambda
\let\X=\Xi

\def\nn{\nonumber} \def\bd{\begin{document}} \def\ed{\end{document}}
\def\ds{\documentstyle} \let\fr=\frac \let\bl=\bigl \let\br=\bigr
\let\Br=\Bigr \let\Bl=\Bigl
\let\bm=\bibitem
\let\na=\nabla
\let\pa=\partial \let\ov=\overline

\def\ft#1#2{{\textstyle{{\scriptstyle #1}\over {\scriptstyle #2}}}}
\def\fft#1#2{{#1 \over #2}}
\def\del{\partial}
\def\sst#1{{\scriptscriptstyle #1}}
 \def\oneone{\rlap 1\mkern4mu{\rm l}}
\def\ie{{\it i.e.\ }}
\def\via{{\it via}}
\def\semi{{\ltimes}}
\def\str{{\rm str}}
\def\tr{{\rm tr}}
\def\Dm{{{D_{\sst{max}}}}}
\def\vac{ \left | 0 \right \rangle }
\def\kvac{ \left | k \right \rangle }

\def\sp{\; \; \;}

\def\bol{ \left | B (p^+) \right \rangle}
\def\bo1{ \left | B^0 (p^+) \right \rangle}

\def\bolt{ \left | B (p^+) \right \rangle_{\t}}

\def\boxl{ \left | B (x^-) \right \rangle}

\def\<{ \langle }
\def\>{ \rangle }

\renewcommand{\floatpagefraction}{0.6}
\renewcommand{\textfraction}{0.2}

\newcommand\ca{\mathcal{A}}
\newcommand\vp{\varphi}
\newcommand\bbone{\ensuremath{\mathbbm{1}}}

\newcommand{\eq}[1]{\begin{equation}#1\end{equation}}
\newcommand{\spl}[1]{\begin{split}#1\end{split}}
\newcommand{\al}[1]{\begin{align}#1\end{align}}
\newcommand{\subeq}[1]{\begin{subequations}#1\end{subequations}}

\newcommand{\arXividhepth}[1]{\href{http://arxiv.org/abs/#1}arXiv:{\tt #1} [hep-th]}
\newcommand{\arXividother}[2]{\href{http://arxiv.org/abs/#1}arXiv:{\tt #1} [#2]}

\newcommand{\bg}[1]{\hat{#1}}
\newcommand{\wj}{\widetilde{J}}
\newcommand{\reo}{\mathrm{Re}~\!\omega}
\newcommand{\imo}{\mathrm{Im}~\!\omega}
\newcommand{\ads}{AdS_4}
\newcommand{\lcal}{\mathcal{L}}
\newcommand{\mcal}{\mathcal{M}}
\newcommand{\ccal}{\mathcal{C}}
\newcommand{\ncal}{\mathcal{N}}
\newcommand{\boxedeq}[1]{
\begin{equation}
\fbox{
\rule[0.7cm]{0pt}{0pt}
$#1$
\rule[-0.45cm]{0pt}{0pt}
}
\end{equation}
}

\def\d{\text{d}}

\def\slashchar#1{\setbox0=\hbox{$#1$}           
\dimen0=\wd0                                 
\setbox1=\hbox{/} \dimen1=\wd1               
\ifdim\dimen0>\dimen1                        
\rlap{\hbox to \dimen0{\hfil/\hfil}}      
#1                                        
\else                                        
\rlap{\hbox to \dimen1{\hfil$#1$\hfil}}   
/                                         
\fi}

\def\Re           {{\rm Re\hskip0.1em}}
\def\Im           {{\rm Im\hskip0.1em}}

\newcommand{\E}{\text{\tiny E}}

\renewcommand{\thefootnote}{\arabic{footnote}}

\title{ Asymptotic symmetries and geometry on the boundary in the first order formalism}

\author{Yegor Korovin\footnote{Postdoctoral Researcher of the Fund for Scientific Research - FNRS Belgium}}
\affiliation{Universit{\'e} Libre de Bruxelles and International Solvay Institutes,\\ C.P. 231, 1050 Brussels, Belgium}

\emailAdd{Yegor.Korovin@ulb.ac.be}

\abstract{Proper understanding of the geometry on the boundary of a spacetime is a critical step on the way to extending holography to spaces with non-AdS asymptotics. In general the boundary cannot be described in terms of the Riemannian geometry and the first order formalism is more appropriate as we show. We analyze the asymptotic symmetries in the first order formalism for large classes of theories on AdS, Lifshitz or flat space. In all cases the asymptotic symmetry algebra is realized on the first order variables as a gauged symmetry algebra. First order formalism geometrizes and simplifies the analysis. We apply our framework to the issue of scale versus conformal invariance in AdS/CFT and obtain new perspective on the structure of asymptotic expansions for AdS and flat spaces.}

\begin{document} 
	\maketitle
	\flushbottom


\section{Introduction}

The boundary of an asymptotically locally AdS (AlAdS) space can be described using Riemannian geometry. Most attempts to extend the AdS/CFT dictionary to non-AlAdS spaces lead to a realization that the geometry on the boundary is typically much less rigid than in the Riemannian case. For Schr\"odinger and Lifshitz spacetimes one is lead to consider torsional Newton-Cartan geometry (TNC) or twistless TNC (TTNC) \cite{Christensen:2013rfa, Hartong:2014oma}. At null infinity of asymptotically flat spaces the boundary is naturally described in terms of Carrollian geometry \cite{Duval:2014uva,Duval:2014lpa}. Both cases share the same feature: the connection on the boundary is not canonically fixed in terms of some underlying structure like metric in the Riemannian case. This clash between Riemannian geometry in the bulk and non-Riemannian geometry on the boundary suggests that the geometry on the boundary should be described using independent frame fields and spin-connection and possibly some other geometric entities. Investigating this idea is the main purpose of this paper.

Let us pause and recall what is the role of the background metric and the stress-energy tensor in a (quantum) field theory. The stress-energy tensor is a conserved current if the theory is coupled in a diffeomorphism invariant way to the background metric. Moreover, conserved currents associated to global spacetime symmetries can be constructed from the stress-energy tensor. For Poincar\'e invariant theory the spin current is given by $J^{\m\n\r} = x^{[\n}T^{\r]\m}$. In a conformal field theory (CFT) the conserved currents are given by $J^{\m} = \xi_{\n}T^{\m\n}$, where $\xi$ is any conformal Killing vector.

What is the underlying reason for the fact, that all the currents can be constructed from the stress-energy tensor alone? The reason is that relativistic field theories can be coupled to the background geometry through the metric, possibly involving covariant derivatives and non-minimal couplings to the curvature. Global spacetime symmetries arise then as symmetries of the background metric, i.e. there exist vector fields solving the Killing (or conformal Killing) equation. The notion of a spacetime symmetry can be formulated using the single object - the background metric. Obviously this is the simplest situation possible. If the geometry is non-Riemannian or is characterized by more than one geometric entity (e.g. covariant derivative is constructed using an independent affine connection), then there may be no such object as the stress-energy tensor, from which all other currents can be derived. Instead there will be independent currents which couple directly to the additional independent geometric data.

Let us recall the story for Poincar\'e invariant field theories in some detail. Spinor representations of the Lorentz group cannot be obtained from representations of $GL(n,\mathbb{R})$, hence one can't couple them covariantly to the background metric. General matter fields can be coupled to the background geometry through the introduction of the frame field $e_{i}^a$ and spin connection $\o_{i}^{ab}$ (latin indices from the beginning of the alphabet refer to the tangent space, whereas those from the middle of the alphabet are world indices). The classical action $S$ is then a functional of the frame field and spin connection and the variations
\be
 T^i_a = \frac{1}{|e|}\frac{\delta S}{\delta e^a_i(x)}, \qquad S^i_{ab}  = \frac{2}{|e|}\frac{\delta S}{\delta \w^{ab}_i(x)}
\ee
define the energy-momentum tensor $T^i_a$ and the spin current $S^i_{ab}$. Note that defined this way $T^i_a$ is not the symmetric Belinfante-Rosenfeld stress-energy tensor which would couple to symmetric metric. To get the symmetric stress-energy tensor we should assume that the spin connection is torsionless, i.e. 
\be
\o_{i}^{ab}(e) = 2 e^{j[a}\partial_{[i}e_{j]}^{b]} - e^{j[a}e^{b]k} e_{i c}\partial_{j}e_k^c,
\ee
where our antisymmetrization convention is such that $2X_{[a}Y_{b]} = X_a Y_b - X_b Y_a$. Then it is clear that the variation with respect to the spin connection is not independent from the variation with respect to the frame fields, in fact\footnote{Here is a quick derivation of this formula. In the absence of torsion $d e^a +\o^a{}_b\wedge e^b =0 $. Variation of it gives $D_{[i}\delta e_{j]}^a + (\delta \o^{ab}_{[i}) e_{j]b}=0$. Next transform tangent space index to a world index $e_{k a}D_{[i}\delta e_{j]}^a + (\delta \o^{ab}_{[i}) e_{j]b} e_{k a}=0$. From the last formula one obtains \eqref{connectionvar} by a standard trick.}
\be
\label{connectionvar}
2 e_j^a e_k^b \delta \o_{i ab} = (D_{[i}\delta e_{j]}^a) e_{ka} - (D_{[j}\delta e_{k]}^a) e_{ia}+(D_{[k}\delta e_{i]}^a) e_{ja},
\ee
where $D_i$ is the Lorentz covariant derivative acting by
\be
D_i \delta e_j^a = \partial_i \delta e_j^a +\o_i^{ab}\delta e_{jb}.
\ee
The total variation of the action is
\begin{align}
\delta S &= \int |e| \Big( T_a^i \delta e_i^a + \frac{1}{2} S^i_{ab} \delta \o_i ^{ab} \Big) \nn \\
&= \int |e| \Big( T_a^i \delta e_i^a + \frac{1}{4} S^{ijk} ((D_{[i}\delta e_{j]}^a) e_{ka} - (D_{[j}\delta e_{k]}^a) e_{ia}+(D_{[k}\delta e_{i]}^a ) e_{ja} \Big)  \nn \\ 
&= \int |e| \Big( T_a^i \delta e_i^a + \frac{1}{4} (S^{ijk} -S^{kij} + S^{jki}) (D_{[i}\delta e_{j]}^a) e_{ka}\Big).
\end{align}
Here $S^{ijk}= S_{ab}^i e^{j a} e^{k b}$ is antisymmetric in the last two indices. Note that we have used the inverse frame field here. Now in the absence of torsion we can replace in the above formula the Lorentz covariant derivative $D$ by covariant derivative $\nabla$ (i.e. the one including the Levi-Cevita affine connection) and after integration by parts we get
\be
\delta S = \int |e| \Big[ T_a^i + \frac{1}{2} \nabla_j (S^{ij}{}_a -S_a{}^{ij} - S^{ji}{}_a) \Big]\delta e_{i}^a.
\ee
The combination in the square bracket is exactly the Belinfante-Rosenfeld tensor. It is this energy-momentum tensor which is symmetric and which couples to metric. Note that the improvement terms involving spin current do not modify the translation charges. From the symmetric energy-momentum tensor one can construct the angular momentum current as $J^{ijk} = x^{[j}T^{k]i}$. It is this current which produces the charge generating Lorentz rotations.

The main lesson here is that there is single object describing all of the geometry. In this case it is the frame field, or alternatively the metric. The operator which couples to it is the energy momentum tensor (of momentum current). Since all the spacetimes symmetries are realised on the metric, the corresponding currents can be derived from the energy momentum tensor. This is the simplest situation possible.

In a more general situation it will not be possible to realize all geometric symmetries using single object. The standard procedure for describing geometry consists of gauging certain algebra by introducing a gauge field for every generator and then one may reduce the number of independent gauge fields by imposing curvature constraints. This is exactly how local Poincar\'e symmetry is realized on the frame field without the need to treat the spin connection as an independent dynamical variable. For other algebras one may not be able to reduce the number of independent gauge fields or one may be able to do it only partially. Even more generally, one may introduce additional symmetries or dynamical matter fields which would allow to express the gauge fields associated to spacetimes symmetries in terms of other dynamical fields.

As a simple example of the phenomenon described above we will consider the case of asymptotically AdS itself in the next section. However in order to produce a non-Riemannian geometry on the boundary we will go beyond the ordinary GR-like setup and will not assume that the spin connection on the boundary is determined in terms of the frame field. Such setup is easily realized in theories with high curvature corrections (or non-minimal coupling to curvature and torsion) in the bulk. In fact we will not assume any field equations in the bulk. The only assumption we make is that the bulk theory can be described using the frame field and the spin connection (plus possible other matter). Such setup will allow us to capture the universal part of the geometry induced on the boundary of AdS, Lifshitz or flat spaces. We will identify the bulk dual of the conformal currents. More precisely, we will show that the frame field and the spin connection in the bulk encode the sources for all the conformal currents on the boundary. Therefore from the space-time symmetry perspective the first order formalism for bulk gravity is more natural than the metric language\footnote{The role of torsion in AdS/CFT was investigated in \cite{Banados:2006fe, Klemm:2007yu, Petkou:2010ve, Blagojevic:2013bu, Cvetkovic:2017fxa}. In particular in \cite{Banados:2006fe} it is pointed out that the independent spin connection gives rise to an independent spin current. However in these works additional assumptions are made, which do not allow to see the full conformal algebra.}. In particular, the first order formulation will clarify the origin of special conformal invariance in AdS/CFT and explain the mechanism behind the known holographic examples of scale but not conformally invariant theories.

Some of our results in section \ref{AdS} are known or can be obtained also in the metric formulation. However for many questions the first order formulation is more transparent and leads to result in a more direct way. Furthermore, for some theories the metric formulation is not available or is not natural. For example higher spin theories are typically formulated in terms of (generalized) frame fields and spin connection.

After dealing with the AdS space we turn to the Lifshitz case in section \ref{Lifshitz} and to flat space in section \ref{Flatspace}. Here the geometry on the boundary does not admit a description in terms of Riemannian geometry and the power of the first order formalism becomes obvious.

One of the main entries in the AdS/CFT dictionary is given by the notion of holographic reconstruction \cite{deHaro:2000vlm}, which relates deformations and states in a CFT to geometries in the bulk. This requires complete knowledge of independent sources on the boundary of CFT. Thankfully, for GR with negative cosmological constant the crucial result in this direction was obtained by Fefferman and Graham \cite{Fefferman:2007rka} and carries the status of a theorem. There is no such a powerful result for flat spaces. One of the main results of this paper will be a proposal for the complete set of sources at the null infinity (see also \cite{Hartong:2015usd} for the $3d$ case).

Let us outline the general strategy we follow in individual examples. Our analysis does not rely on field equations in any way, we only assume that the gravitational theory admits spaces with prescribed asymptotics as solutions and that it can be formulated in the first order formalism, i.e. using frame field and spin-connection for the Lorentz group. In particular, we do not impose zero torsion constraint, and thus treat frame field and spin connection as independent variables. Hence our analysis applies to a much larger class of gravitational theories than GR (with cosmological constant). 

As usual, the analysis of asymptotic symmetries requires some guesswork and assumptions. Let us however outline the steps we follow in all the examples considered below.
\begin{enumerate}
\item Begin by fixing the gauge. We fix the $E_r^{\hat{A}}$ and $\O_r^{\hat{A}\hat{B}}$ components of the frame field and spin connection. Here $r$ corresponds to a radial direction (which will be specified in every case separately) and $\hat{A}$ stands for all tangent space indices. $E_i^{\hat{A}}$ and $\O_i^{\hat{A}\hat{B}}$ components can potentially play the role of gauge fields on the boundary.
\item Make assumptions on the leading fall off behaviour of $E_i^{\hat{A}}$. This specifies the class of spaces we are dealing with.
\item Solve for residual gauge transformation which preserve the gauge and fall off conditions.
\item Impose fall off conditions on the $\O_i^{\hat{A} \hat{B}}$ components of the spin connection in a way which would be consistent with the zero torsion constraint. We do not however solve the zero torsion constraint.
\item Identify the (gauged) algebra of residual gauge transformations.
\end{enumerate}

In the bulk the frame and the spin connection transform under covariant general coordinate transformations (parametrised by $\Xi^{\m}$ or $\X^{\hA} = E_{\m}^{\hA} \X^{\m}$) and local Lorentz rotation ($\L^{\hA \hB}$) according to
\al{
\delta E_{\m}^{\hA} &= \pa_{\m} \X^{\hA} - \L^{\hA \hB}E_{\m \hB} + \X_{\hB} \O_{\m}^{\hA \hB} + \X^{\n} T_{\n \m}^{\hA}, \\
\delta \O_{\m}^{\hA \hB} &= \pa_{\m}\L^{\hA \hB} + 2 \O_{\m}^{\hat{C} [\hA} \L^{\hB]}{}_{\hat{C}}+ \X^{\n} R_{\n \m}^{\hA \hB},
}
where $T_{\m\n}^{\hA}$ is the torsion tensor and $R_{\m\n}^{\hA\hB}$ is the Riemann tensor (see Appendix \ref{GaugeBasics}).

\section{Conformal currents in $AdS$ holography}
\label{AdS}

\subsection{Gauged conformal algebra on the boundary of AdS}
Let us turn for a moment to general CFTs. One should distinguish between the conformal currents and the stress-energy tensor. The former ones correspond to global space-time symmetries, whereas the on-shell conservation of the latter follows from the diffeomorphism invariance. 
Conformal currents and the stress-energy tensor are of course related in a simple way in a CFT. Conformal currents are obtained from\footnote{We reserve the latin indices from the middle of the alphabet for the world indices in CFT, whereas greek indices are kept for the bulk. Hatted indicies refer to the tangent space.}
\be
J_{(A)}^{i} = \xi_{(A) j} T^{ij},
\ee
where $\xi_{(A) i}$ is the Killing vector and $(A)$ labels the generators of the conformal group $SO(d,2)$. Usually one couples a CFT to an external metric $g_{ij}$ which sources the stress-energy tensor $T^{ij}$. In AdS/CFT it is well understood that the bulk metric $g_{\m\n}$ is dual to the boundary stress-energy tensor. However the conformal currents are as good operators in CFT as the stress-energy tensor itself. What is the corresponding bulk field? Since we are speaking about space-time symmetries we do not want to introduce any additional gauge fields in the bulk. Instead the bulk field should be related to the geometry in some way. We propose that the bulk fields which are dual to conformal currents are the frame field $E^{\hA}_{\m}$ and the spin connection $\O^{\hA \hB}_{\m}$. The goal of this section is to confirm this claim by analyzing the asymptotic symmetries in the first order formalism (before imposing any constraint which would express the spin connection in terms of the frame field).  To see it in more detail let us introduce the sources for the conformal currents in the CFT:
\be
\int d^dx \Big( P^{i}_{\ha} e^{\ha}_{i} + M^{i}_{\ha \hb} \o^{\ha \hb}_{i} + D^{i}b_{i} + K^{i}_{\ha} f^{\ha}_{i} \Big),
\ee
where $P^{i}_a$ are the currents for translations ($d$ of them: $a=1,\ldots,d$), $M^{i}_{ab}$ is the current for Lorentz rotations, $D^{i}$ stands for the dilatation current and $K^{i}_a$ are the special conformal currents. $e_i^{\ha}$ is the frame field to which the CFT couples, $\o_i^{\ha \hb}$ is the spin connection, $b_i$ is the source for the dilatation current and $f_i^{\ha}$ is the source for the special conformal current. The correlation functions of this operators are obtained by differentiating the generating functionals with respect to corresponding sources.

Under conformal transformations the sources in CFT transform according to (see Appendix \ref{GaugeBasics} for details) \begin{align}
\label{confvar}
\delta e^{\ha}_{i} &=\pa_i\x^{\ha}+ \x_{\hb} \o_i^{\ha \hb}- \l^{\ha \hb} e_{i \hb} + \x^{\ha} b_i- \l_D e^{\ha}_{i} +\x^j(2\pa_{[j}e_{i]}^{\ha} - 2 e_{[j}^{\ha} b_{i]} + 2 \o_{[j}^{\ha \hb} e_{i]\hb}), \\
\delta \o_{i}^{\ha \hb} &= \pa_{i}\l^{\ha \hb} + 2 \o_{i \hc}{}^{[\ha}\l^{\hb]\hc} - 4 \l_K^{[\ha}e^{\hb]}_{i} + 4 \x^{[\ha}f^{\hb]}_{i} +\x^j(2\pa_{[j}\o_{i]}^{\ha \hb} + 2 \o_{[j}^{\hc \ha} \o^{\hb}_{i]\hc} + 8 f_{[j}^{[\ha} e_{i]}^{\hb]}),\\
\delta b_{i} &= \pa_{i} \l_D + 2 \l_K^{\ha} e_{i \ha} - 2 \x^{\ha} f_{i\ha} + \x^j(2\pa_{[j}b_{i]} + 4 e_{[j}^{\ha} f_{i]a}),\\
\label{confvarf}
\delta f^{\ha}_{i} &= \pa_{i}\l_K^{\ha} - b_{i} \l_K^{\ha} + \o_{i}^{\ha \hb} \l_{K \hb} - \l^{\ha \hb} f_{i \hb} + \l_D f^{\ha}_{i} + \x^j(2\pa_{[j}f_{i]}^{\ha} + 2 f_{[j}^{\ha} b_{i]} + 2 \o_{[j}^{\ha \hb} f_{i]\hb}).
\end{align}

Now consider the frame field $E^{\hA}_{\m}$ and the spin connection $\O^{\hA \hB}_{\m}$ in the bulk. The conformal boundary of an AlAdS space defines the radial direction $r$ orthogonal to it. Thus at least near the  conformal boundary the fields can be decomposed with respect to the radial direction as
\be
E^{\hat{A}}_{\m} \rightarrow (E^{\hat{a}}_{i}, E^{\hat{r}}_{i};E_r^{\hat{a}},E_r^{\hat{r}}), \qquad \O^{\hat{A}\hat{B}}_{\m} \rightarrow (\O^{\hat{a}\hat{b}}_{i}, \O^{\hat{r}\hat{a}}_{i};\O_r^{\hat{a}\hat{b}},\O_r^{\hat{r}\hat{a}}).
\ee
We decompose analogously the parameters for the bulk coordinate transformations $\X^{\hA}$ and local Lorentz transformations $\L^{\hA \hB}$. In the bulk we have $d+1$ local translations (equivalent to covariant general coordinate transformations if the torsion vanishes) and $d(d+1)/2$ local Lorentz rotations which in total gives $(d+1)(d+2)/2$ parameters. We can use this gauge freedom to fix certain components of the fields. We impose
\be
\label{gaugefix}
E_r^{\hat{r}} =-1, \qquad E_r^{\hat{a}} = 0, \qquad \O_r^{\hat{r}\hat{a}}=0,\qquad \O_r^{\hat{a}\hat{b}}=0.
\ee
The remaining components have a potential to become a set of gauge fields on the boundary. Conditions \eqref{gaugefix} fix completely the gauge in the bulk, meaning that there are no residual gauge transformation parameters which would depend on all bulk coordinates. It is worth pointing out that our gauge \eqref{gaugefix} differs from the commonly used Fefferman-Graham gauge (with $g_{ri}=0$), which requires $E_i^{\hat{r}} =0$. As we shall see, keeping $E_i^{\hat{r}} \neq 0$ is necessary to exhibit the full conformal algebra acting on the boundary sources. 

The naive expectation is that the $\O^{\ha \hb}_{i}$ would provide the source for the Lorentz currents $\o^{\ha \hb}_i$ on the boundary, $E^{\hr}_{i}$ should correspond to the dilatation source $b_i$ and certain combinations of $E^{\ha}_{i}$ and $\O^{\hr \ha}_{i}$ are related to $e^{\ha}_{i}$ and $f^{\ha}_{i}$. The remainder of this section confirms this expectation.

The gauge above is preserved by residual gauge transformations which satisfy
\begin{align}
\label{gaugecondeq}
0&=\pa_r \X^{\hat{r}} - \X^i T_{ri}^{\hat{r}}, \\
\label{gaugecondeq1}
0&=\pa_r \X^{\hat{a}} -\L^{\hat{a}\hat{r}}E_{r \hat{r}}- \X^i T_{ri}^{\hat{a}}, \\
\label{gaugecondeq12}
0&=\pa_r\L^{\hat{r}\hat{a}} + \X^i R_{ir}^{\hr \ha} = \pa_r\L^{\hat{r}\hat{a}} -\X^i \pa_r \O_{i}^{\hr \ha},\\
0&=\pa_r\L^{\hat{a}\hat{b}} + \X^i R_{ir}^{\ha \hb} = \pa_r\L^{\hat{a}\hat{b}} - \X^i \pa_r \O_{i}^{\ha \hb}.
\label{gaugecondeq2}
\end{align}
Solutions to this system of equations define the analog of Penrose-Brown-Henneaux (PBH) transformations. 

One can solve equations \eqref{gaugecondeq}-\eqref{gaugecondeq2} systematically as follows. \eqref{gaugecondeq} and \eqref{gaugecondeq2} can be integrated immediately assuming that torsion vanishes to relevant order. To obtain a decoupled equation for $\X^i$ we act with the radial derivative on \eqref{gaugecondeq1} and use \eqref{gaugecondeq12}. The decoupled equation reads
\be
\label{decoupledXi}
E_i^{\ha} \pa_r^2 \X^i + 2 \pa_r E_i^{\ha} \pa_r \X^i=0.
\ee
The general solution for $\X^i$ involves two integration functions\footnote{The equation \eqref{decoupledXi} is solved by 
	\be
	\label{ressol}
	\X^j = \a^j(x) + \int^r \exp(-2\int^{r'} M_i^j(r'',x)dr'')dr' \b^i(x),
	\ee
	where the components of the matrix
	$M$ are given by $M_i^j(r'',x)= E_i^{\ha} \pa_{r''} E^j_{\ha}$ and $\a^j$ and $\b^j$ are arbitrary integration functions. Here we have to assume that the frame field is invertible at least asymptotically.} $\X^i = \a^i(x) + e^{-2r} \b^i(x) + \ldots$. $\L^{\hr \ha}$ is then obtained from \eqref{gaugecondeq1}. Finally the allowed gauge transformations are parametrized by 
\begin{align}
\X^{\ha}(r,x) &= e^r \Xi_{(0)}^{\ha}(x) + e^{-r} \Xi_{(2)}^{\ha} (x) + \ldots, \\
\X^{\hr}(r,x) &= \l_D(x) + \ldots, \\
\L^{\hat{r}\hat{a}}(r,x) &= e^r \Xi_{(0)}^{\ha}(x) - e^{-r} \Xi_{(2)}^{\ha} (x) + \ldots,\\
\L^{\hat{a} \hat{b}}(r,x)&= \l^{\hat{a} \hat{b}}(x) + \ldots,
\end{align}
where $\Xi_{(0)}^{\ha}(x)$, $\Xi_{(2)}^{\ha}(x)$, $\l_D(x)$ and $\l^{\hat{a} \hat{b}}(x)$ are arbitrary (integration) functions on the boundary.

Our next task is to determine the fall off behaviour of individual components of the frame field and the spin connection. We want to keep the analysis as general as possible and will not assume particular field equations. We want however to be compatible with the general relativity (GR), and therefore we assume fall off behaviour which does not immediately contradict the one in GR. In particular we assume that certain components of the torsion vanish asymptotically. The no-torsion constraint reads in the differential form language
\be
T^{\hA} = dE^{\hA} + \O^{\hA}{}_{\hB}\wedge E^{\hB}=0,
\ee
or in components
\begin{align}
\label{torsioni}
T_{ri}^{\hat{r}} &= \pa_r E_i^{\hat{r}}, \\
T_{ri}^{\hat{a}} &= \pa_r E_i^{\hat{a}} - \O_i^{\hat{a}\hat{r}} E_{r \hat{r}}, \\
T_{ij}^{\hat{r}} &= 2 \pa_{[i}E_{j]}^{\hat{r}} + 2 \O_{[i}^{\hat{r} \hat{a}} E_{j]\hat{a}}, \\
T_{ij}^{\hat{a}} &=2 \pa_{[i}E_{j]}^{\hat{a}} + 2 \O_{[i}^{ \hat{a} \hat{r}} E_{j]\hat{r}} + 2 \O_{[i}^{\hat{a} \hat{b}} E_{j]\hat{b}}.
\label{torsionf}
\end{align}

In $AdS$ space the $E_i^{\hat{a}}$ component of the frame field goes as $e^r$ near the boundary (plus subleading terms of course), which we locate at $r = \infty$. Solving the zero torsion constraints for the spin-connection we would find that $\O_i^{\hat{r}\hat{a}}$ scales in the same way as $E_i^{\hat{a}}$. Thus we write
\al{
E_i^{\hat{a}} &=  e_{i}^{\hat{a}}(x) e^r+ e_{(2)i}^{\ha}(x) e^{-r} +\ldots,\\
E_i^{\hat{r}} &=  b_{i}(x)+\ldots,\\
\O_i^{\hr \ha} &= e_{i}^{\hat{a}}(x) e^r - e_{(2)i}^{\ha}(x) e^{-r} +\ldots \\
\O_i^{\ha \hb} &= \o_i^{\ha \hb}(x) + \ldots
}
Similar asymptotic expansion in $AdS_3$ was adopted in \cite{Grumiller:2016pqb}.
We allowed for the terms in the near-boundary expansions which could transform under residual gauge transformations.

The remaining components transform according to:
\begin{align}
\label{transi}
\delta E_i^{\hat{r}} &= \pa_i \X^{\hr} - \L^{\hat{r}\hat{a}} E_{i \hat{a}} + \O_i^{\hr \ha} \X_{\ha} + \X^r T_{ri}^{\hr} + \X^j T_{ji}^{\hr}, \\
\delta E_i^{\hat{a}} &= \pa_r \X^{\hat{a}}  - \L^{\hat{a} \hat{r}} E_{i \hat{r}} - \L^{\hat{a} \hat{b}} E_{i \hat{b}} + \O_i^{\ha \hr} \X_{\hr}+ \O_i^{\ha \hb} \X_{\hb}  + \X^r T_{ri}^{\ha} + \X^j T_{ji}^{\ha}, \\
\delta \O_i^{\hat{r} \hat{a}} &= \pa_i \L^{\hat{r} \hat{a}} +2 \O_{i\hat{b}}{}^{[\hat{r}} \L^{\hat{a}] \hat{b}} +\X^r R_{ri}^{\hr \ha}+\X^j R_{ji}^{\hr \ha},\\
\delta \O_i^{\hat{a} \hat{b}} &= \pa_i \L^{\hat{a} \hat{b}} +2 \O_{i\hat{r}}{}^{[\hat{a}} \L^{\hat{b}] \hat{r}} +2 \O_{i\hat{c}}{}^{[\hat{a}} \L^{\hat{b}] \hat{c}}+\X^r R_{ri}^{\ha \hb}+\X^j R_{ji}^{\ha \hb}.
\label{transf}
\end{align}
 
It is convenient to define a field $F_i^a$ by \cite{Banados:2006fe}
\be
F_i^{\ha} = \frac{1}{2}(E_i^{\ha} - \O_i^{\hr \ha}),
\ee
so that its expansion near the boundary starts with $F_i^{\ha} = e_{(2)i}^{\ha} e^{-r} + \ldots$ and it transforms according to
\al{
\delta F_i^{\ha} = &\frac{1}{2}\pa_i(\X^{\ha} - \L^{\hr \ha}) - \L^{\ha \hb} F_{i \hb} - \frac{1}{2}\L^{\ha \hr} E_{i \hr} + \frac{1}{2} \O_i^{\ha \hr} \X_{\hr} + \frac{1}{2}\O_i^{\ha \hb} (\X_{\hb} - \L_{\hr \hb}) \nn \\ +& \X^r \pa_r F_i^{\ha} + \frac{1}{2}\X^r \O_i^{\ha \hr} + \X^j \O_{[j}^{\ha \hr}E_{i]\hr} + \X^j (2\pa_{[j}F_{i]}^{\ha} + 2 \O_{[j}^{\ha \hb} F_{i]}^{\hb}).
}
Keeping only the leading order terms in the transformations of $E_i^{\hat{r}}$, $E_i^{\hat{a}}$, $F_i^{\hat{a}}$ and $\O_i^{\hat{a} \hat{b}}$ we arrive at
\al{
\delta b_i &= \pa_i \l_D + 2 \X_{(2)}^{\ha} e_{i\ha} - 2 \X_{(0)}^{\ha} e_{(2)i\ha} + \X_{(0)}^j(2 \pa_{[j}e_{i]}^{\hr} + 4 e_{[j}^{\ha} e_{(2)i]\ha}), \label{dilatat} \\
\delta e_i^{\ha} &= \pa_i \X^{\ha}_{(0)} \!+\! \X^{\ha}_{(0)} b_i \!-\! e_i^{\ha} \l_D - \l^{\ha \hb} e_{i \hb} + \o_i^{\ha \hb} \X_{(0) \hb} + \X^j_{(0)}(2 \pa_{[j}e_{i]}^{\ha} \!-\! 2 e_{[j}^{\ha} b_{i]} + 2 \o_{[j}^{\ha \hb} e_{i]\hb}),\\
\delta \o_i^{\ha \hb} &=\! \pa_i \l^{\ha \hb} \!+\! 2 \o_{i\hc}{}^{[\ha} \l^{\hb] \hc} \!+\! 4 e_i^{[\ha} \X_{(2)}^{\hb]}\!-\! 4 \X_{(0)}^{[\ha} e_{(2)i}^{\hb]} \!+\! \X_{(0)}^j(2 \pa_{[j}\o_{i]}^{\ha \hb} \!+\! 2 \o_{[j}^{\hc \ha} \o_{i]}^{\hb}{}_{\hc} + 8 e_{(2)[j}^{[\ha} e_{i]}^{\hb]}),\\
\delta e_{(2)i}^{\ha} &\!=\! \pa_i \X^{\ha}_{(2)} \!-\! \l^{\ha \hb} e_{(2) i \hb} + \o_i^{\ha \hb} \X_{(2) \hb} + \l_D e_{(2)i}^{\ha}-b_i \X^{\ha}_{(2)} + 
2\X^j_{(0)}(\pa_{[j}e_{(2)i]}^{\ha} + e_{(2)[j}^{\ha} b_{i]} + \o_{[j}^{\ha \hb} e_{(2)i]\hb}).
\label{specconf}
}
We recognize the gauged conformal algebra upon identification
\be
\X^{\ha}_{(0)} = \x^{\ha}, \qquad \X^{\ha}_{(2)} = \l_K^{\ha}, \qquad e_{(2)i}^{\ha} = f_i^{\ha}.
\ee
This is the main result of this section. The asymptotic symmetry algebra is realized on the first order variables as the gauged algebra. Using \eqref{gaugetransform} one can read off the commutation relations of the corresponding symmetry algebra directly from the transformation properties of the frame field and the spin connection. There is no need to compute the commutator of two transformations preserving the gauge. The algebra reads
\al{
[M_{ab},M_{cd}] &= 4 \eta_{[a[c} M_{d]b]}, \qquad [P_a, M_{bc}] = 2 \eta_{a[b}P_{c]}, \qquad [K_a, M_{bc}] = 2 \eta_{a[b}K_{c]}, \nn \\
[P_a,K_b] &= 2(\eta_{ab}D + M_{ab}), \qquad [D,P_a]=P_a, \qquad [D,K_a] = - K_a.
}
Let us emphasize that in the spirit of the AdS/CFT correspondence we allowed for arbitrary sources on the boundary. If we fix the sources to particular values, the asymptotic symmetry algebra will reduce, e.g. if one fixes the metric on the boundary of $AdS_3$ to be flat, one obtains Virasoro algebra \cite{Brown:1986nw}. 

\subsection{Relation to the metric approach}
Usually in AdS/CFT the geometry on the boundary is described in terms of the metric. How does our description above relates to the metric one? Similarly as for the Poincar\'{e} algebra, to get a describtion in terms of Riemannian geometry one imposes curvature constraints which allow to solve algebraically for the spin connection $\o_i^{\ha\hb}$ and the field $f_i^{\ha}$. Specifically, the no-torsion constraint $R_{ij}(P^{\ha})=0$ gives an algebraic equation for the spin connection, whereas $R_{ij}(M^{\ha \hb})=0$ allows us to solve for $f_i^{\ha}$ algebraically. The solution is
\be
f_i^{\ha} = -\frac{1}{2(d-2)} \Big(R_i^{\ha} - \frac{1}{2(d-1)}e_i^{\ha}R\Big),
\ee
where $R_i^{\ha}$/$R$ is the Ricci tensor/scalar associated with $\o_i^{\ha\hb}[e_i^{\ha}]$. If we now plug this solution into the near boundary expansion and assume (or gauge fix) $E_i^{\hr}=0$, we obtain the expansion for the metric
\be
ds^2 =dr^2+ e^{2r}(e_i^{\ha} + e^{-2r} f_i^{\ha} +\ldots)(e_{j\ha} + e^{-2r} f_{j\ha} +\ldots) = dr^2 + e^{2r}(g_{(0)ij}+e^{-2r} g_{(2)ij} + \ldots),
\ee
where \cite{Skenderis:1999nb}
\be
g_{(2)ij} = 2 e_i^{\ha} f_{j \ha} = -\frac{1}{(d-2)} \Big(R_{(0)ij} - \frac{1}{2(d-1)} R_{(0)} g_{(0)ij} \Big).
\ee
It is known that this expression for the $g_{(2)ij}$ coefficient is universal, i.e. it is true in most theories of gravity (see \cite{Imbimbo:1999bj} for a cohomological argument and \cite{Aksteiner:2015uxw} for an explicit computation). From our perspective this result follows from the algebraic constraint $R_{ij}(M^{\ha \hb})=0$ and the relation between the $g_{(2)ij}$ coefficient and the gauge field for special conformal transformations $f_i^{\ha}$.

\subsection{Scale vs. Conformal invariance in holography}
Our analysis geometrizes individual conformal transformation on the bondary. In first order formalism the special conformal transformations on the boundary are induced by special local Lorentz rotations in the bulk - the ones which mix radial direction with the boundary ones. As a small application we now clarify the issue of scale vs. conformal invariance in holography, which has been addressed by Nakayama in a series of papers (see \cite{Nakayama:2013is} for a review). 

The example in \cite{Nakayama:2012sn} is provided by foliation preserving gravity (full diffeomorphism invariance is broken by the terms involving extrinsic curvature in the action). 
Foliation preserving diffeomorphisms provide local translations and rotations on the boundary. Scale invariance arises from translations in radial direction. However the local rotations mixing boundary and radial directions are broken. As we have seen above these are exactly the bulk transformations which induce special conformal transformation on the boundary. This breaking leads to $R^2$-term in the trace anomaly for four-dimensional field theory.

Another example was given in \cite{Nakayama:2009qu}, where it was argued that backgrounds involving massive vector field $A = \a dr$ lead to breaking of special conformal invariance while preserving Poincare and scaling invariance. From the point of view of asymptotic symmetries realised in the first order formalism it is clear why this example works. It obviously preserves translations and scaling symmetries as well as local rotations in the plane along the boundary. However the rotation mixing radial with boundary directions would break Lorentz invariance on the boundary and thus is not allowed on such a background.

\subsection{Conserved currents and the improved stress-energy tensor}

Let us for completeness discuss certain relations which are implied by the spacetime symmetries in a conformal field theory. In the context of the AdS/CFT correspondence these become Ward identities for correlation functions. The difference with respect to the usual treatment is that instead of the metric we have individual source for every conformal current. Our discussion follows closely that in \cite{deWit:1981vgr} (see also \cite{Didenko:2012vh} for the corresponding treatment involving couplings to higher spin currents).

Expanding the conformal gauge fields around their flat space value we have the action (in this section we drop the hats on the tangent space indices for convenience)
\be
S = S_{\text{matter}} +\int d^dx \Big( P^{i}_a e^{a}_{i} + M^{i}_{ab} \o^{ab}_{i} + D^{i}b_{i} + K^{i}_a f^{a}_{i} \Big),
\ee
where $S_{\text{matter}}$ stands for the action in flat space. Now we vary the total action. Assuming that it is conformally invariant and that the field equations are satisfied in flat space we get
\be
\int d^dx  \Big( P^{i}_a \delta e^{a}_{i} + M^{i}_{ab} \delta \o^{ab}_{i} + D^{i} \delta b_{i} + K^{i}_a \delta f^{a}_{i} \Big) =0.
\ee
Using the transformation rule for the gauge fields \eqref{confvar}-\eqref{confvarf} we arrive at
\al{
0&= \pa_i P_a^i, \\
0&= \pa_i M_{ab}^i + \frac{1}{2}(P_{ab} - P_{ba}) , \\
0&= \pa_i D^i + P_i^i , \\
0&= \pa_i K_a^i - 2 M_{ia}^i -D_a .
}
Not all these currents are conserved. However it is possible to define $x$-dependent combinations, which are:
\al{
0&= \pa_i(M_{ab}^i + \frac{1}{2}(x_a P^i_{b} - x_b P_a^i)) , \\
0&= \pa_i(D^i + x^a P_a^i)) , \\
0&= \pa_i(K_{a}^i - 2 M^i_{ab} x^b - D^i x_a + P^i_{b}(\frac{1}{2}x^2\delta_{ab} - x_a x_b)) .
}
So far we assumed that all the sources are independent, and hence the currents are independent too. If however the constraints $R_{ij}(P^{\ha})=0$ and $R_{ij}(M^{\ha \hb})=0$ are imposed the number of independent currents is reduced. The dependence of $\o_i^{ab}$ and $f_i^a$ on $e_i^a$ leads to improvement terms for the stress-energy tensor $P_a^i$. The field $b_i$ should decouple \cite{deWit:1981vgr}. At the end the improved stress-energy tensor is conserved, symmetric and traceless. The individual conformal currents are obtained by contracting the improved stress-energy tensor with the conformal Killing vectors. Thus we made the contact with the conventional treatments of conformal currents.

\section{Asymptotically Lifshitz spaces}
\label{Lifshitz}
Let us start with several comments. 
For asymptotically Lifshitz case the geometry on the boundary depends to some extent on the matter content of the bulk theory. In many well-studied examples there is either an additional gauge $U(1)$ symmetry in the bulk (Einstein-Maxwell-dilation model) or some special matter (St\"uckelberg field). Additional symmetry allows to make spin connection and boost connection composite fields. However this reduction does rely crucially on the special properties of the matter content in the bulk. Even though these cases are well motivated phenomenologically, one would definitely benefit from a more model-independent understanding of the boundary geometry. This is one of the main goals of this section.

Lifshitz space is not a solution of Einstein-Hilbert gravity. One needs matter fields \cite{Kachru:2008yh,Taylor:2008tg} or modifications of GR \cite{AyonBeato:2009nh, Cai:2009ac} to support it (see \cite{Taylor:2015glc} for a review). Clearly, the dual field theory interpretation and even the asymptotic symmetry algebra depends on particular theory under consideration. As in the previous section, we are not restricting to particular theory, instead we will try to find the largest possible symmetry algebra which can be realized on the frame field and spin-connection. The advantage of this is that we will be able to clearly see what boundary conditions should be allowed by a theory in order to realize given symmetry algebra. On the other hand we will not be sensitive to additional gauge fields or to twists involving internal and space-time gauge fields.

The metric of the Lifshitz space is given by
\be
ds^2_{Lif} = dr^2 - e^{2 z r} dt^2 + e^{2r} dx_a dx^a.
\ee
For concreteness we will assume in the following that the dynamical exponent $z$ lies in the interval $1 \leq z \leq 2$. Extension to $z < 1$ can be done along the same lines. It will be also clear that for integer $z$, e.g. for $z=2$ the analysis has to be done separately due to additional symmetries.

The main difference with respect to AdS is that there is no natural non-degenerate metric induced on the boundary (i.e. as $r \rightarrow \infty$). There is however a metric on the spatial sections of the boundary parametrized by $x^a$. 

Again we start by fixing the gauge:
\be
\label{Lifgauge}
E_r^{\hat{r}}=-1, \qquad E_r^{\hat{t}} = 0 = E_r^{\hat{a}}, \qquad \O_r^{\hat{A}\hat{B}}=0,
\ee
where we decompose the tangent space index $\hat{A} = (\hat{r}, \hat{t}, \hat{a})$. In addition we assume the following fall-off conditions (see \cite{Ross:2011gu,Hartong:2014oma}):
\be
\label{Liffalloff}
E_i^{\hat{t}} \sim e^{z r} e_{(0)i}^{\hat{t}}, \qquad E_i^{\hat{a}} \sim e^r e_{(0)i}^{\hat{a}}, \qquad E_i^{\hat{r}} \sim e^0 b_i.
\ee
The components of the torsion tensor are:
\begin{align}
T_{ri}^{\hat{r}} &= \pa_r E_i^{\hat{r}}, \\
\label{Trit}
T_{ri}^{\hat{t}} &= \pa_r E_i^{\hat{t}}-\O_i^{\hat{t}\hat{r}}E_{r \hat{r}}, \\
\label{Tria}
T_{ri}^{\hat{a}} &= \pa_r E_i^{\hat{a}}-\O_i^{\hat{a}\hat{r}}E_{r \hat{r}},\\
\label{Tijr}
T_{ij}^{\hat{r}} &= 2\pa_{[i} E_{j]}^{\hat{r}} + 2 \O_{[i}^{\hat{r} \hat{t}} E_{j] \hat{t}} + 2 \O_{[i}^{\hat{r} \hat{a}} E_{j] \hat{a}},\\
\label{Tijt}
T_{ij}^{\hat{t}} &= 2\pa_{[i} E_{j]}^{\hat{t}} + 2 \O_{[i}^{\hat{t} \hat{r}} E_{j] \hat{r}} + 2 \O_{[i}^{\hat{t} \hat{a}} E_{j] \hat{a}},\\
\label{Tija}
T_{ij}^{\hat{a}} &= 2\pa_{[i} E_{j]}^{\hat{a}} + 2 \O_{[i}^{\hat{a} \hat{t}} E_{j] \hat{t}} + 2 \O_{[i}^{\hat{a} \hat{r}} E_{j] \hat{r}}+ 2 \O_{[i}^{\hat{a} \hat{b}} E_{j] \hat{b}}.
\end{align}
We want the fall-off of the spin-connection to be consistent with the zero-torsion constraint, i.e.
\be
\O_i^{\hat{a}\hat{b}} \sim \op(e^{(2z-2)r}), \qquad \O_i^{\hat{a}\hat{t}} \sim \op(e^{(z-1)r}), \qquad \O_i^{\hat{a}\hat{r}} \sim \op(e^r), \qquad  \O_i^{\hat{t}\hat{r}} \sim \op(e^{zr}).
\ee
Setting the relevant components of the torsion to zero asymptotically would imply
\begin{align}
\O_i^{\hat{a}\hat{b}} &= - e_{(0)}^{j[\ha} e_{(0)}^{\hb] k} e_{(0) i \hat{t}}\pa_j e_{(0)k}^{\hat{t}} e^{(2z-2)r} + \ldots , \\
\O_i^{\hat{a}\hat{t}} &= \Big(e_{(0)}^{j\ha} \pa_{[i}e_{(0)j]}^{\hat{t}}- e_{(0)}^{k[\ha} e_{(0)}^{\hat{t}]j} e_{(0) i \hat{t}}\pa_k e_{(0)j}^{\hat{t}} +\frac{z}{2}e_{(0)k}^{\hat{r}}e_{(0)}^{k\ha}e_{(0) i}^{\hat{t}} \Big)e^{(z-1)r} + \ldots.
\end{align}
Also for the moment we are not specifying the scaling of the subleading terms in the expansions. Our goal will be to determine the most general asymptotic algebra consistent with these conditions.

The gauge \eqref{Lifgauge} is preserved by local translations and Lorentz rotations satisfying
\begin{align}
\delta E_r^{\hat{r}} &= \partial_r \X^{\hr} -\X^i T_{ri}^{\hr}=0, \\
\label{dErt}
\delta E_r^{\hat{t}} &= \partial_r \X^{\hat{t}} -\L^{\hat{t} \hat{r}} E_{r\hat{r}}  -\X^i T_{ri}^{\hat{t}}=0,\\
\label{dEra}
\delta E_r^{\hat{a}} &= \partial_r \X^{\ha} -\L^{\hat{a} \hat{r}} E_{r\hat{r}}  -\X^i T_{ri}^{\hat{a}}=0,\\
\delta \O_r^{\hat{A} \hat{B}} &= \pa_r\L^{\hat{A} \hat{B}} - \X^i \pa_r\O_i^{\hat{A} \hat{B}} =0.
\end{align}
Residual transformations are parametrized by
\al{
\X^{\hr} &= \X_{(0)}^{\hr} + \ldots, \\
\X^i &= \X^i_{(0)} + e^{-2r}\X^i_{(2)}(x)+ e^{-2zr}\X^i_{(2z)}(x), \\
\L^{\ha \hb} &= \l_{(0)}^{\ha\hb} \!+\! \X^i_{(0)} (\O_i^{\ha\hb}\!-\!\o_i^{\ha\hb}) \!+\!\frac{z\!-\!1}{z\!-\!2}\X^i_{(2)}\O_{(0)i}^{\ha\hb}e^{(2z-4)r}+ (1-z) \X^i_{(2z)}\O_{(0)i}^{\ha\hb}e^{-2r} +\ldots, \\
\label{latresidual}
\L^{\ha \hat{t}} &= \l_{(0)}^{\ha\hat{t}} \!+\! \X^i_{(0)} (\O_i^{\ha\hat{t}}\!-\!\o_i^{\ha\hat{t}}) +\frac{z\!-\!1}{z\!-\!3}\X^i_{(2)}\O_{(0)i}^{\ha\hat{t}}e^{(z-3)r}+ \frac{1\!-\!z}{z\!+\!1} \X^i_{(2z)}\O_{(0)i}^{\ha\hat{t}}e^{-(1+z)r}+\ldots,
}
with $\X^i_{(2)}$ and $\X^i_{(2z)}$ constrained to satisfy
\be
\X^i_{(2)} e_{(0)i}^{\hat{t}}=0, \qquad \X^i_{(2z)} e_{(0)i}^{\hat{a}}=0.
\ee
Note that these constraints are not Lorentz-covariant. Modulo these constraints, $\X_{(0)}^{\hr}$, $\X^i_{(0)}$, $\l_{(0)}^{\ha\hb}$, $\l_{(0)}^{\ha\hat{t}}$, $\X^i_{(2)}$ and $\X^i_{(2z)}$ are arbitrary (integration) functions depending on the boundary coordinates only. Note that we can alternatively write
\al{
\X^{\ha} &= e^r \X_{(0)}^{\ha} + e^{-r} \X_{(2)}^{\ha} + \ldots, \qquad
\X^{\hat{t}} = e^{zr} \X_{(0)}^{\hat{t}} + e^{-zr} \X_{(2z)}^{\hat{t}} + \ldots,
}
where all denoted coefficients are arbitrary integration functions. We also assume
\al{
\O_i^{\ha\hb} &= \O_{(0)i}^{\ha\hb}e^{(2z-2)r} + \o_i^{\ha\hb} + \ldots, \\
\O_i^{\ha\hat{t}} &= \O_{(0)i}^{\ha\hat{t}}e^{(z-1)r} + \o_i^{\ha\hat{t}} + \ldots.
}
The gauge conditions \eqref{dErt} and \eqref{dEra} imply
\be
\L^{\hat{r} \hat{a}} = e^r \X^{\ha}_{(0)}(x) - e^{- r}\X^{\ha}_{(2)}(x)+\ldots, \qquad \L^{\hat{r} \hat{t}} =z e^{zr} \X^{\hat{t}}_{(0)}(x) -z e^{- zr}\X^{\hat{t}}_{(2z)}(x)+\ldots.
\ee
We will also allow the terms at corresponding orders in the expansion of the frame field and the spin connection, which now take the form
\begin{align}
E_i^{\hat{r}} &= b_i(x), \\
\label{Oitrfalloff}
E_i^{\hat{t}} &= e^{z r} e_{(0)i}^{\hat{t}} +e^{-z r} e_{(2)i}^{\hat{t}} + \ldots, \qquad \O_i^{\hat{t} \hat{r}}= -z e^{z r} e_{(0)i}^{\hat{t}} + z e^{-z r} e_{(2)i}^{\hat{t}} + \ldots, \\
\label{Oiarfalloff}
E_i^{\hat{a}} &= e^{ r} e_{(0)i}^{\hat{a}} + e^{ -r} e_{(2)i}^{\hat{a}} + \ldots, \qquad \O_i^{\hat{a} \hat{r}}= -e^{ r} e_{(0)i}^{\hat{a}} + e^{ -r} e_{(2)i}^{\hat{a}} + \ldots.
\end{align}

Next we have to check if all these residual transformations preserve the fall-off conditions \eqref{Liffalloff}.
Individual components transform as
\begin{align}
\label{dEir}
\delta E_i^{\hat{r}} &= \pa_{i} \X^{\hat{r}} \!-\! \L^{\hat{r} \hat{t}}  E_{i\hat{t}} \!-\! \L^{\hat{r} \hat{a}} E_{i\hat{a}} +\X_{\hat{t}} \O_i^{\hr \hat{t}} +\X_{\hat{a}} \O_i^{\hr \hat{a}} +2\X^j(\pa_{[j} E_{i]}^{\hat{r}} +  \O_{[j}^{\hat{r} \hat{t}} E_{i] \hat{t}} +  \O_{[j}^{\hat{r} \hat{a}} E_{i] \hat{a}}), \\
\label{dEit}
\delta E_i^{\hat{t}} &= \pa_{i} \X^{\hat{t}} \!-\! \L^{\hat{t} \hat{r}} E_{i\hat{r}} \!-\! \L^{\hat{t} \hat{a}} E_{i\hat{a}}+\X_{\hat{r}} \O_i^{ \hat{t} \hr } +\X_{\hat{a}} \O_i^{\hat{t}\ha } +2\X^j(\pa_{[j} E_{i]}^{\hat{t}} +  \O_{[j}^{\hat{t} \hat{r}} E_{i] \hat{r}} + \O_{[j}^{\hat{t} \hat{a}} E_{i] \hat{a}}) , \\
\label{dEia}
\delta E_i^{\hat{a}} &=  \pa_{i} \X^{\hat{a}}  - \L^{\hat{a} \hat{r}} E_{i\hat{r}} - \L^{\hat{a}\hat{t} } E_{i\hat{t}} - \L^{\hat{a}\hat{b}} E_{i\hat{b}} +\X_{\hat{r}} \O_i^{ \hat{a} \hr } +\X_{\hat{t}} \O_i^{\hat{a}\hat{t} }+\X_{\hat{b}} \O_i^{\hat{a}\hb } \\ &\qquad\qquad\qquad\qquad\qquad+\X^j(2\pa_{[j} E_{i]}^{\hat{a}} + 2 \O_{[j}^{\hat{a} \hat{t}} E_{i] \hat{t}} + 2 \O_{[j}^{\hat{a} \hat{r}} E_{i] \hat{r}}+ 2 \O_{[j}^{\hat{a} \hat{b}} E_{i] \hat{b}}), \\
\delta \O_i^{\hat{r} \hat{a}} &= \pa_i \L^{\hat{r} \hat{a}} +2 \O_{i\hat{t}}{}^{[\hat{r}} \L^{\hat{a}]\hat{t}} +2 \O_{i\hat{b}}{}^{[\hat{r}} \L^{\hat{a}]\hat{b}} +\X^r \pa_r \O_i^{\hat{r} \hat{a}} \nn \\  &\qquad\qquad\qquad\qquad\qquad+\X^j(2 \pa_{[j}\O_{i]}^{\hr \ha} + 2 \O_{[j}^{\hat{t} \hr} \O_{i]\hat{t}}^{\ha} + 2 \O_{[j}^{\hat{b} \hr} \O_{i]\hat{b}}^{\ha}),\\
\delta \O_i^{\hat{a} \hat{b}} &=\pa_i \L^{ \hat{a} \hat{b}}+2 \O_{i\hat{r}}{}^{[\hat{a}} \L^{\hat{b}]\hat{r}} +2 \O_{i\hat{t}}{}^{[\hat{a}} \L^{\hat{b}]\hat{t}} +2 \O_{i\hat{c}}{}^{[\hat{a}} \L^{\hat{b}]\hat{c}}+ \X^r \pa_r \O_i^{\hat{a} \hat{b}} \nn \\  &\qquad\qquad\qquad\qquad\qquad+ \X^j(2 \pa_{[j}\O_{i]}^{\ha \hb} + 2 \O_{[j}^{\hat{t} \ha} \O_{i]\hat{t}}^{\hb} + 2 \O_{[j}^{\hat{r} \ha} \O_{i]\hat{r}}^{\hb}+ 2 \O_{[j}^{\hat{c} \ha} \O_{i]\hat{c}}^{\hb}),\\
\delta \O_i^{\hat{r} \hat{t}} &= \pa_i \L^{\hat{r} \hat{t}} +2 \O_{i\hat{a}}{}^{[\hat{r}} \L^{\hat{t}]\hat{a}} +\X^r \pa_r \O_i^{\hat{r} \hat{t}} + \X^j(2 \pa_{[j}\O_{i]}^{\hr \hat{t}} + 2 \O_{[j}^{\hat{a} \hr} \O_{i]\hat{a}}^{\hat{t}} ),\\
\delta \O_i^{\hat{t}\hat{a}} &=  \pa_i \L^{\hat{t}\hat{a}}\!+\!2 \O_{i\hat{r}}{}^{[\hat{t}} \L^{\hat{a}]\hat{r}} \!+\!2 \O_{i\hat{b}}{}^{[\hat{t}} \L^{\hat{a}]\hat{b}}\! +\!\X^r \pa_r \O_i^{\hat{t}\hat{a}}  + 2\X^j( \pa_{[j}\O_{i]}^{\hat{t} \ha} +  \O_{[j}^{\hat{r} \hat{t}} \O_{i]\hat{r}}^{\ha} +  \O_{[j}^{\hat{b} \hat{t}} \O_{i]\hat{b}}^{\ha}).
\end{align}
Now we have to check if all the integration functions preserve the fall-off behaviour. It is easy to see that $\X_{(2)}^{\ha}$ spoils the behaviour \eqref{Oiarfalloff} unless it is set to zero. Beyond that we have to set
\be
\l_{(0)}^{\ha\hat{t}}=\X_{(0)}^j \o_j^{\ha \hat{t}}
\ee
in order to preserve the fall-off behaviour of $E_i^{\ha}$. Then the entire bulk rotation $\L^{\ha\hat{t}}$ is completely fixed by \eqref{latresidual} without allowing for any integration function. Moreover, now we have to make sure that the algebra closes. In other words the commutators of other transformations are not allowed to produce terms in $\L^{\ha\hat{t}}$ other than those given by \eqref{latresidual}. This however can happen, since the bulk rotations parametrized by $\L^{\hr \ha}$ and $\L^{\hat{t}\hr}$ do commute into $\L^{\ha\hat{t}}$. Specifically, consequent transformation by $\X_{(0)}^{\ha}$ and $\X_{(2z)}^{\hat{t}}$ would produce a $\L^{\ha\hat{t}}$ transformation which is not allowed. Thus we have to set either $\X_{(0)}^{\ha}$ or $\X_{(2z)}^{\hat{t}}$ to zero. The former choice would effectively break the translational invariance on the boundary and the resulting symmetry algebra would be a direct product of $\mathfrak{sl}(2)$ (parametrized by time translations $H$, dilatation $D$ and special conformal generator $K$) and spatial rotations. We prefer to keep the full translational invariance on the boundary and set impose instead $\X_{(2z )}^{\hat{t}}=0$. Then asymptotically we remain with
\begin{align}
\delta e^{\hat{a}}_{(0)i} &=\pa_{i} \X_{(0)}^{\ha} +\X_{(0)}^{\ha} b_i -\X_{(0)\hr}e^{\hat{a}}_{(0)i} - \l^{\ha \hb} e_{(0)i \hb} + \X_{(0)\hb} \o^{\ha \hb}_{i} \\& \qquad+ \X^j_{(0)}(2 \pa_{[j}e_{(0)i]}^{\ha} -2  e^{\hat{a}}_{(0)[j}b_{i]} + 2 \o^{\ha \hb}_{[j} e_{(0)i]\hb}), \\
\delta e^{\hat{t}}_{(0)i} &= \pa_{i} \X_{(0)}^{\hat{t}} +z(\X_{(0)}^{\hat{t}} b_i -\X_{(0)\hr}e^{\hat{t}}_{(0)i})+ \X^j_{(0)}(2 \pa_{[j}e_{(0)i]}^{\hat{t}} -2z  e^{\hat{t}}_{(0)[j}b_{i]} ), \\
\delta b_{i} &= \pa_{i} \X_{(0)}^{\hr} +2\X^j_{(0)} \pa_{[j}b_{i]} ,\\
\delta \o_{i}^{\hat{a}\hat{b}} &= \pa_{i}\l^{\hat{a}\hat{b}} + 2 \o_{i \hat{c}}{}^{[\hat{a}}\l^{\hat{b}]\hat{c}}  + 2\X^j_{(0)}(\pa_{[j}\o_{i]}^{\hat{a} \hb} +  \o_{[j}^{\hc \ha} \o^{\hb}_{i]\hc}).
\end{align}
This is the gauged Lifshitz algebra consisting of space ($P^{\ha}$) and time ($H$) translations, space rotations ($M^{\ha\hb}$) and dilatations ($D$). The non-vanishing commutators are
\begin{align}
[D,P^{\ha}] &= P^{\ha},\\
[D,H] &= z H,\\
[M^{\ha\hb},P^{\hc}] &= 2\delta^{\hc [\hb}P^{\ha]},\\
[M^{\ha\hb},M^{\hc\hat{d}}] &= 4\delta^{[\ha [\hc}M^{\hat{d}]\hb]}.
\end{align}

Let us finish with a couple of comments regarding the special case of $z=2$. To see why it is so special consider the bulk boost $\L^{\ha \hat{t}}$ given in \eqref{latresidual}. It has a term of order $e^{-r}$ involving an integration function $\X_{(2)}^i$. It is exactly of the right order to be able to rotate $e_{(0)i}^{\hat{t}}$ into $e_{(0)i}^{\hat{a}}$, i.e. to act as a boost. Note that this can happen only for $z$ satisfying $z-3 = 1-z$, i.e. $z=2$. Thus this new algebra would involve non-relativistic boosts for $z=2$. Moreover the space translations $\Xi_{(0)}^{\ha}$ and timelike special conformal transformation $\X_{(4)}^{\hat{t}}$ commute into boosts. Thus the Lifshitz algebra gets enhanced by the boosts and timelike special conformal transformation. This is exactly the structure of the Schr{\"o}dinger algebra.

\section{Asymptotically flat spaces}
\label{Flatspace}

\subsection{Spatial infinity}

Many formulas remain the same as in the $AdS$ case. We adopt the gauge \eqref{gaugefix}, which is preserved only by local translations and rotations satisfying \eqref{gaugecondeq}-\eqref{gaugecondeq2}. The no-torsion constraint takes the same form as in the $AdS$ case, see equations \eqref{torsioni}-\eqref{torsionf}.

We expand the frame field and the spin connection near the boundary in power series\footnote{More generally there can be a logarithmic term in the expansion of $E_i^{\ha}$. We adopt this simplified fall-off behaviour. It will suffice to exhibit the Poincar\'e algebra as the asymptotic symmetry algebra.}:
\al{
E_i^{\ha} &= r e_{(0)i}^{\ha} + e_{(2)i}^{\ha} + \ldots,\\
E_i^{\hat{r}} &=  h_{i}(x)+\ldots,\\
\O_i^{\hr \ha} &= e_{(0)i}^{\hat{a}}(x) +\ldots, \\
\O_i^{\ha \hb} &= \o_i^{\ha \hb}(x) + \ldots.
}
The residual local translations and Lorentz rotations are parametrized by\footnote{One can use \eqref{decoupledXi} and \eqref{ressol} to find the integration functions in the expansion of $\Xi^{\ha}$.}
\al{
\X^{\ha} &= r \X_{(0)}^{\ha}+\X_{(2)}^{\ha} +\ldots, \\
\L^{\hr \ha} &= \X_{(0)}^{\ha} + \ldots, \\
\X^{\hr} &= \X_{(0)}^{\hr} + \ldots,\\
\L^{\ha \hb} &= \l^{\ha \hb} + \ldots,
}
where $\X_{(0)}^{\ha}$, $\X_{(2)}^{\ha}$, $\X_{(0)}^{\hr}$ and $\l^{\ha \hb}$ are arbitrary (integration) functions on the boundary.

Transformations of individual components are given by \eqref{transi}-\eqref{transf}. If we define a new variable 
\be
F_i^{\ha} = E_i^{\ha}-r\O_i^{\hr \ha} \sim  e_{(2)i}^{\ha} + \ldots
\ee
then the transformations of the leading terms in the expansions of $E_i^{\hr}$, $E_i^{\ha}$, $F_i^{\ha}$ and $\O_i^{\ha\hb}$ are
\al{
\label{Palgi}
\delta h_i &= \pa_i\X_{(0)}^{\hr} + e_{(0)i}^{\hat{a}} \X_{(2)\ha} - e_{(2)i}^{\hat{a}} \X_{(0)\ha}+\X_{(0)}^j (2\pa_{[j}e_{(0)i]}^{\hr} + 2 e_{(0)[j}^{\ha}e_{(2)i]\ha}),\\
\delta e_{(0)i}^{\hat{a}} &= \pa_i\X_{(0)}^{\ha} -  \l^{\ha \hb} e_{(0)i\hb} + \o_i^{\ha \hb} \X_{(0)\hb} + \X_{(0)}^j (2\pa_{[j}e_{(0)i]}^{\ha} + 2 \o_{[j}^{\ha \hb} e_{(0)i]\hb}),\\
\delta e_{(2)i}^{\hat{a}} &= \pa_i\X_{(2)}^{\ha} -  \l^{\ha \hb} e_{(2)i\hb} + \o_i^{\ha \hb} \X_{(2)\hb} + \X_{(0)}^{\ha} h_i - \X_{(0)}^{\hr}e_{(0)i}^{\hat{a}} \nn \\& \qquad\qquad\qquad\qquad\qquad+\X_{(0)}^j (2 \pa_{[j}e_{(2)i}^{\ha} - e_{(0)[j}^{\ha} h_{i]} + 2 \o_{[j}^{\ha \hb} e_{(2)i]\hb}),\\
\delta  \o_i^{\ha \hb} &= \pa_i \l^{\ha \hb} + 2 \o_{i\hc}{}^{[\ha} \l^{\hb]\hc} + \X_{(0)}^j (2\pa_{[j}\o_{i]}^{\ha\hb} + 2 \o_{[j}^{\hc \ha} \o_{i]\hc}^{\hb}).
\label{Palgf}
}
How are we to interpret this algebra? Let us set to zero the curvature/torsion terms. Then upon identifying $e_{(0)i}^{\ha} = \o_i^{\ha}$ and $e_{(2)i}^{\ha} = e_{i}^{\ha}$, we recognize the gauged Poincar\'{e} algebra, where $h_i$ is the gauge field for "radial" translations (would be time translations near future/past infinity $i^+$/$i^-$), and $\o_i^{\ha}$ is the gauge field for the boosts. The unusual feature of \eqref{Palgi}-\eqref{Palgf} is that the curvature terms come multiplied with $\X_{(0)}^j$, which now corresponds to the boost parameter and not the translation parameter. This property is clearly inherited from the bulk. It just happens that leading terms in the expansion of translations in the bulk correspond to boosts on the boundary.

In the metric formulation one usually adopts the hyperbolic slicing near spatial infinity. The commonly used Beig-Schmidt expansion of the metric near spatial infinity takes the following form:
\be
ds^2 = dr^2 + r^2 \Big(g_{(0)ij}(x) + \frac{1}{r}g_{(1)ij}(x) + \ldots \Big)dx^i dx^j.
\ee
In general relativity $g_{(0)ij}(x)$ must satisfy Einstein equation with negative cosmological constant
\be
\label{bconstraint}
R_{ij}[g_{(0)}] = (d-1)g_{(0)ij}.
\ee
In four spacetime dimensions, $g_{(0)ij}(x)$ describes locally a unit-normalized three-dimensional hyperboloid. From the first order formalism perspective however, we identify
\be
g_{(0)ij} = \o_i^{\ha} \o_{j\ha}, \qquad g_{(1)ij} = 2\o_i^{\ha} e_{j\ha}.
\ee
Thus the metric is not given by the square of the frame field, but rather by the square of the boost gauge field. In fact it is the $g_{(2)ij}$ coefficient which has a chance to be the square of the frame field (if there is no other independent boundary data appearing at that order)! Actually there is no reason why $e_i^{\ha}$ should even be invertible (as a frame field on "hyperboloid").

In fact our point of view now provides new perspective on the constraint \eqref{bconstraint}. As in $AdS$ case we may want to impose some curvature constraints and reduce the number of independent gauge fields. In $AdS$ it automatically allowed us to solve for $g_{(2)ij}$ in terms of the Schouten tensor associated with $g_{(0)ij}$. What happens when we impose similar constraint at spatial infinity? First of all note that the bulk torsion $T_{ij}^{\ha}$ near spatial infinity reduces to
\be
T_{ij}^{\ha} = d \o^{\ha} + \o^{\ha \hb}\wedge \o_{\hb} + \op\Big(\frac{1}{r}\Big),
\ee
i.e. asymptotically it becomes the curvature associated with boosts! Imposing no-torsion asymptotically allows us to solve for the spin connection $\o^{\ha \hb}$ algebraically, however not in terms of the frame field, but in terms of the boost connection. In addition we have to assume that the boost frame field $\o_i^{\ha}$ is invertible.

If we now in addition impose the vanishing of the Lorentz curvature
\be
R^{\ha\hb} = d \o^{\ha \hb} + \o_{\hc}{}^{\ha} \wedge \o^{\hb \hc} - \o^{\ha}\wedge \o^{\hb}=0,
\ee
we realize that it becomes a true differential constraint on $\o^{\ha}$. In fact it is exactly equivalent to \eqref{bconstraint}. Thus the differential constraint on $g_{(0)ij}$ in the Beig-Schmidt expansion at $i^0$ has the same algebraic origin as the algebraic solution for $g_{(2)ij}$ in the Fefferman-Graham expansion in $AdS$!

Interestingly, in the metric formulation the additional constraint $g_{(0)ij} \propto g_{(1)ij}$ is frequently imposed (see e.g. \cite{Mann:2006bd, Mann:2008ay, Compere:2011db}). From the first order formalism perspective this constraint makes frame field and the boost connection proportional to each other!

Clearly our analysis of the geometry suggests novel notion of the covariance at $i^0$. It is tempting to speculate that it can pave the road to a novel approach to holographic renormalization and bulk reconstruction of asymptotically flat spaces.

\subsection{Null infinity}

In asymptotically flat case the boundary at $\cal{I}^+$ is null, and we decompose the tangent space index as $\hat{A} = (\hat{r},\hat{u},\hat{a})$, where we think of $\hat{u}$ direction as the null-tangent to the $\cal{I}^+$. The metric on the tangent space is off-diagonal in the $(\hat{r},\hat{u})$-plane $\eta_{\hat{r}\hat{u}}=-1$ and diagonal in remaining directions: $\eta_{\hat{a}\hat{b}}=\delta_{\hat{a}\hat{b}}$ We are going to work in null coordinates, so that the Minkowski background itself is\footnote{To obtain this expression of the metric from the familiar form
\be
ds^2 = - dt^2 + dx^2 + \sum_a d\tilde{x}_a^2
\ee
we change coordinates according to $r = t-x$, $\tilde{x}_i = r x_i$ and $u = t+x - (t-x)\sum_i x_i^2$.}

\be
ds^2 = - dr du + r^{2} dx_a dx^a.
\ee
For pure Minkowski space $x^a$ are coordinates on the plane. Often other coordinates are used, so that $x^a$ describe a sphere. For our local analysis it will not be important if we slice the null infinity with spheres or with planes since we are going to allow for arbitrary metric on the slice . We impose the gauge
\be
\label{flatgauge}
E_r^{\hat{u}}=1, \qquad E_r^{\hat{r}}=0=E_r^{\hat{a}}, \qquad \O_r^{\hA\hB}=0
\ee
and assume the following fall-off behaviour:
\be
E_i^{\hat{u}} \sim \op(r), \qquad E_i^{\hat{r}} \sim \op(1), \qquad E_i^{\hat{a}} \sim \op({r}).
\ee
Similar fall-off conditions were adopted in \cite{Hartong:2015usd}.
The gauge \eqref{flatgauge} breaks anisotropy on the boundary and is preserved by $\X$ and $\L$ satisfying
\begin{align}
0&=\pa_r \X^{\hr}, \\
0&=\pa_r \X^{\hu} - \L^{\hu \hr} E_{r\hr}, \\
0&=\pa_r \X^{\ha} - \L^{\ha \hr} E_{r\hr}, \\
0&=\pa_r \L^{\hr\hu} - \X^i \pa_r \O^{\hr \hu}_i,\\
0&=\pa_r \L^{\hr\ha} - \X^i \pa_r \O^{\hr \ha}_i,\\
0&=\pa_r \L^{\hu\ha} - \X^i \pa_r \O^{\hu \ha}_i,\\
0&=\pa_r \L^{\ha\hb} - \X^i \pa_r \O^{\ha \hb}_i,
\end{align}
where we neglected the torsion terms by assuming that they vanish to relevant order. These conditions are solved by
\al{
\X^{\hr} &= \X^{\hr}_{(0)},\\
\X^{\hu} &= r \X^{\hu}_{(0)} +\X^{\hu}_{(1)},\\
\X^{\ha} &= r \X^{\ha}_{(0)} +\X^{\ha}_{(1)},\\
\L^{\hr\hu} &= \X^{\hu}_{(0)} + \op(r^{-2}), \\
\L^{\hr\ha} &= \X^{\ha}_{(0)} + \op(r^{-2}), \\
\L^{\ha\hb} &= \l^{\ha \hb}_{} + \op(r^{-1}), \\
\L^{\hu\ha} &= \l^{\ha}_{(0)} + \op(r^{-1}).
}
To determine the fall off behaviour of the spin connection we consider the components of the torsion:
\begin{align}
T_{ri}^{\hat{u}} &= \pa_r E_i^{\hat{u}}, \\
\label{fTriv}
T_{ri}^{\hat{r}} &= \pa_r E_i^{\hat{r}}-\O_i^{\hat{r} \hu}{}E_{r \hat{u}}, \\
\label{fTria}
T_{ri}^{\hat{a}} &= \pa_r E_i^{\hat{a}}-\O_i^{\hat{a}}{}_{\hat{u}}E_{r}^{\hat{u}},\\
\label{fTiju}
T_{ij}^{\hat{u}} &= 2\pa_{[i} E_{j]}^{\hat{u}} + 2 \O_{[i}^{\hat{u}\hat{r}} E_{j]\hat{r}} + 2 \O_{[i}^{\hat{u}}{}_{\hat{a}} E_{j]}^{\hat{a}},\\
\label{fTijv}
T_{ij}^{\hat{r}} &= 2\pa_{[i} E_{j]}^{\hat{r}} + 2 \O_{[i}^{\hat{r}\hat{u}} E_{j]\hat{u}} + 2 \O_{[i}^{\hat{r}}{}_{\hat{a}} E_{j]}^{\hat{a}},\\
\label{fTija}
T_{ij}^{\hat{a}} &= 2\pa_{[i} E_{j]}^{\hat{a}} + 2 \O_{[i}^{\hat{a}}{}_{\hat{r}} E_{j]}^{\hat{r}} +2 \O_{[i}^{\hat{a}}{}_{\hat{u}} E_{j]}^{\hat{u}}+ 2 \O_{[i}^{\hat{a}}{}_{\hat{b}} E_{j]}^{\hat{b}}.
\end{align}
Note that in these formulas the location of the indices is important.
We expand the fields close to the null-infinity in the manner which is consistent with the zero-torsion constraint:
\al{
E_i^{\hr} &= b_i,\\
E_i^{\hu} &= r h_{(0)i} +h_{(1)i},\\
E_i^{\ha} &= r e^{\ha}_{(0)i} +e^{\ha}_{(1)i},\\
\O_i^{\hr\hu} &= h_{(0)i} + \op(r^{-2}), \\
\O_i^{\hr\ha} &= e^{\ha}_{(0)i} + \op(r^{-2}), \\
\O_i^{\ha\hb} &= \o^{\ha \hb}_{i} + \op(r^{-1}), \\
\O_i^{\hu\ha} &= \o^{\ha}_{(0)i} + \op(r^{-1}).
}

After tedious but straightforward computation we find that the individual components transform in the following way:
\al{
	\delta b_i &= \pa_i \X^{\hr}_{(0)} + \X^{\hu}_{(0)} b_i - \X^{\hr}_{(0)} h_{(0)i} - \X^{\ha}_{(0)}e_{(1)i\ha} + \X^{\ha}_{(1)}e_{(0)i\ha} + \ldots, \\
\delta h_{(0)i} &= \pa_i \X^{\hu}_{(0)} + \X^{\ha}_{(0)} \o_{i\ha} - \l^{\ha}_{(0)} e_{(0)i \ha} +\ldots, \\
\delta e^{\ha}_{(0)i} &= \pa_i \X^{\ha}_{(0)} - \X^{\ha}_{(0)} h_{(0)i} - \l^{\ha\hb}_{(0)} e_{(0)i \hb} + e_{(0)i}^{\ha}\X^{\hu}_{(0)} +\o_i^{\ha\hb} \X_{(0)\hb}  +\ldots, \\
\delta e^{\ha}_{(1)i} &= \pa_i \X^{\ha}_{(1)} \!-\! \X^{\ha}_{(0)} h_{(1)i} \!-\!\l_{(0)}^{\ha}b_i\!-\! \l^{\ha\hb}_{(0)} e_{(1)i \hb} + e_{(0)i}^{\ha}\X^{\hu}_{(1)}+\o_{(0)i}^{\ha} \X_{(0)}^{\hr} +\o_i^{\ha\hb} \X_{(1)\hb} +\ldots, \\
\delta h_{(1)i} &= \pa_i \X^{\hu}_{(1)} - \X^{\hu}_{(0)} h_{(1)i} -\l_{(0)}^{\ha}e_{(1)i\ha} + h_{(0)i} \X^{\hu}_{(1)}+\o_{(0)i}^{\ha} \X_{(1)}^{\ha}  +\ldots,\\
\delta \o^{\ha}_{(0)i} &= \pa_i \l^{\ha}_{(0)} - \X^{\hu}_{(0)} \o^{\ha}_{(0)i} - \l^{\ha\hb}_{(0)} \o_{(0)i \hb} + h_{(0)i} \l^{\ha}_{(0)} +\o_i^{\ha\hb} \l_{(0)\hb} +\ldots, \\
\delta \o^{\ha\hb}_{(0)i} &= \pa_i \l^{\ha\hb}_{(0)} +2\o_{(0)i}^{[\ha} \X_{(0)}^{\hb]}+2 e_{(0)i}^{[\ha} \l_{(0)}^{\hb]} +2\o_{(0)i\hc}{}^{[\ha} \l^{\hb]\hc}_{(0)}  +\ldots,
}
where we have omitted the torsion and curvature terms for the readability purposes. The non-vanishing commutators are
\al{
[P_a, B_b] &= \delta_{ab} P', \qquad [D,P'] = P', \qquad [D, P_a] = P_a, \qquad [K_a, P_b] = \delta_{ab} D + M_{ab}, \\
[P_a, M_{bc}] &= 2 \delta_{a[b}P_{c]}, \qquad [K_a, M_{bc}] = 2 \delta_{a[b}K_{c]}, \qquad [B_a, M_{bc}] = 2 \delta_{a[b}B_{c]}, \\
[P_a, K'] &= B_a, \qquad [P', K_a] = B_a, \qquad [B_a, K_b] =\delta_{ab} K', \\
[D, K_a] &= -K_a, \qquad [D,K'] = -K', \qquad [M_{ab}, M_{cd}] = 4 \delta_{[a[c}M_{c]b]}.
}
We checked explicitly that the Jacobi identities are satisfied.
Note, that $D$ is a Cartan generator. Our notation is summarized in Table \ref{table:1}.

\begin{table}[h!]
\centering
\begin{tabular}{ |c|c|c|c|c|c|c|c| } 
 \hline
 Generator & $P_a$ & $-K_a$ & $M_{ab}$ & $-D$ & $P'$ & $K'$ & $B_a$ \\ 
 \hline
 Gauge field & $e_{(0)i}^{\ha}$ & $\o_{(0)i}^{\ha}$ & $\o_i^{\ha \hb}$ & $h_{(0)i}$ & $b_i$ & $h_{(1)i}$ & $e^{\ha}_{(1)i}$ \\ 
 \hline
 Gauge parameter & $\X_{(0)}^{\ha}$ & $\l_{(0)}^{\ha}$ & $\l^{\ha \hb}$ & $\X_{(0)}^{\hu}$ & $\X_{(0)}^{\hr}$ & $\X_{(1)}^{\hu}$ & $\X_{(1)}^{\ha}$ \\ 
 \hline
\end{tabular}
\caption{Generators, gauge fields and corresponding parameters}
\label{table:1}
\end{table}

In many respects the generator $P'$ looks like another translation $P_a$, $K'$ looks like another $K_a$, whereas the 'boost' $B_a$ looks like another $M_{ab}$. So they might fit into a conformal algebra, however the commutators are not quite those of the conformal algebra in $d$ dimensions. In fact the generators $(D, M_{ab}, P_a, K_a)$ form a $(d-1)$-dimensional conformal algebra (up to a rescaling of $P_a$ and $P'$ by a factor of two) and the set of generators $(P', K', B_a)$ form a normal subalgebra! Thus the resulting algebra is not semisimple. The generator $P'$ is readily identified with the conventional supertranslations. Yet another way to look at this algebra is to note that $P_a$, $P'$, $B_a$ and $M_{ab}$ form a Carroll algebra, with $B_a$ playing the role of the Carroll boosts.

Let us review the standard approach to the geometry on the null-infinity. The lightlike conformal boundary is a Carrollian manifold \cite{Duval:2014uva, Duval:2014lpa}.  Carroll manifolds $C$ are constructed out of Riemannian hypersurfaces $\S$ as $C = \S \times R$. Due to degeneracy of the metric, there is no canonical definition of the connection. Conformal Carroll group of level $N$ is generated by
\be
X = (\o^a_b x^b + \g^a +(\c - 2 \k_b x^b)x^a + \k^a x_b x^b)\pa_a + \Big(\frac{2}{N}(\c - 2 \k_b x^b)u + T(x) \Big)\pa_u.
\ee
The first terms represents conformal transformation on $\S$, whereas $T(x)$ parametrises supertranslations. Carrollian algebra is the semidirect product of conformal algebra on $\S$ and supertranslations. 

In our approach we allow all sources to transform, which leads to important differences. First of all the usual conformal Carrollian algebra is gauged. Moreover, the normal subgroup consists not only of supertranslations $P'$ along the null generators of the boundary, but it posesses larger normal subgroup including additional transformation $K'$ and Carroll boosts $B_a$.

Finally, let us consider the implications of the zero torsion constraints. Setting to zero the leading components of $T_{ij}^{\ha}$,$T_{ij}^{\hu}$ and $T_{ij}^{\hr}$ we obtain
\al{
0&=\pa_{[i}e_{(0)j]}^{\ha} + \o_{[i}^{\ha\hb} e_{(0)j]\hb} + e_{(0)[i}^{\ha} h_{(0)j]}, \\
0&=\pa_{[i}h_{(0)j]} + \o_{[i}^{\ha} e_{(0)j]\ha},\\
0&=\pa_{[i}b_{(0)j]} + e_{(0)[i}^{\ha} e_{(1)j]\ha} + b_{(0)[i}^{\ha} h_{(0)j]},
}
Thus the Lorentz and the special conformal connections become composites of the frame field and the dilatation connection. $e_{(1)i}^{\ha}$ also becomes composite. Furthermore, setting to zero the leading term in $R_{ij}^{\ha \hb}$ we can solve for the special conformal connection $\o_{i}^{\ha}$ in terms of the Schouten tensor on the celestial sphere (similarly as in AdS case) if the dimension of the sphere is bigger than two. For two-dimensional sphere (i.e. four-dimensional flat space) there is no constraint of $\o_{i}^{\ha}$, since the Einstein equation in two dimensions is trivially satisfied.

It would be interesting to work out in detail the source-VEV relation in GR using the first order formalism. We leave it for future work.

\section{Summary and discussion}
We have demonstrated that working in the first order formalism in the bulk, one can avoid the clash between Riemannian geometry in the bulk and non-Riemannian geometry on the boundary. Frame field and the spin connection nicely decompose into a set of boundary gauge fields, which transform under residual local bulk transformations. These residual transformations correspond to gauged boundary spacetime symmetries. The asymptotic symmetry algebra is realized on the frame field and spin connection as a gauged algebra. In asymptotically AdS case it is a gauged conformal algebra. At spatial infinity of the flat space it is a gauged Poincar\'{e} algebra. At null infinity it is a non-semisimple algebra which falls short to be conformal. In fact it is a conformal algebra on the celestial sphere extended by conventional supertranslation, and less conventional 'boosts' $B_a$ and transformation $K'$.

Our analysis applies to a broad class of gravitational theories which extends well beyond GR. The only assumption we make is that the fall-off behaviour of the fields is consistent with the dynamics of the theory. This assumption is satisfied in generic gravities with higher curvature corrections \cite{Aksteiner:2015uxw}.

In the first order formalism the Penrose-Brown-Henneaux transformations (and analogs thereof in flat space) obtain a clear geometrical meaning. In particular it is straightforward to identify individual local transformations in the bulk corresponding to individual conformal transformations on the boundary. This allowed us to shed new light on the question of scale vs. conformal invariance in holography. We believe that our analysis provides a useful perspective on the asymptotic expansions in flat space at spatial and null infinities.

Being so general and entirely geometric, our analysis is blind to the actual dynamics of the bulk theory. Only taking the dynamics into account can one determine the complete set of independent sources on the boundary. Effectively this may reduce a number of independent gauge fields by imposing some zero curvature constraints. Particular theory will tell us, what is the set of fields on the boundary which is needed in order to realize the boundary spacetime symmetries. Our analysis provides the maximal set of gauge fields to which the corresponding conserved currents couple.

\section*{Acknowledgements}
I am grateful to Sergei Kuzenko and Stefan Theisen for insightful discussions on the related problems. It is my pleasure to thank Geoffrey Comp{\`e}re, Jelle Hartong and Zhenya Skvortsov for useful comments on the draft. I am especially grateful to C{\'e}dric Troessaert for numerous invaluable discussions around this work and related topics. The research of Y.
Korovin is supported by the Fund for Scientific Research-FNRS Belgium, grant FC 17085.

\appendix

\section{Gauge transformations for the spacetime symmetry group}
\label{GaugeBasics}

According to standard definitions the gauge fields transform according to
\be
\label{gaugetransform}
\delta(\e)B_{\m}^{\a} = \pa_{\m}\e^{\a} + \e^{\g}B_{\m}^{\b} f_{\b\g}{}^{\a},
\ee
where the individual symmetries are indexed by greek indices from the beginning of the alphabet and $f_{\b\g}{}^{\a}$ are the structure constants of the symmetry algebra. The commutator is given by
\be
[\delta(\e_1), \delta(\e_2)] B_{\m}^{\a} = \pa_{\m}\e_3 + \e_3^{\g} B_{\m}^{\b} f_{\b \g}{}^{\a} = \delta(\e_3)B_{\m}^{\a},
\ee
where
\be
\e_3^{\a} = \e_2^{\g} \e_1^{\b} f_{\b \g}{}^{\a}.
\ee

When we are gauging spacetime symmetries, additional subtleties arise. 
 Knowing the transformation rule under global the Poincar\'{e} (or conformal) symmetry, one can reconstruct the transformation under the gauged group as we describe now (see e.g. \cite{Freedman:2012zz} for more details). Local translations are replaced by general coordinate transformations, in particular they absorb the so-called 'orbital parts' of Lorentz transformations (and of special conformal and scale transformations for the case of gauged conformal algebra). As an example, consider the transformation of the scalar $\phi$ under global Poincar\'{e} transformation:
\be
\delta \phi(x) = (a^{\m} + \l^{\m\n} x_{\n})\pa_{\m}\phi(x).
\ee
Upon gauging the Poincar\'{e} algebra ($a^{\m} \rightarrow \x^{\m}(x)$, $\l^{\m\n} \rightarrow \l^{\m\n}(x)$) we absorb the term $\l^{\m\n} x_{\n}$ into the $\x^{\m}$, so that the action of general coordinate transformation on scalar becomes simply
\be
\label{gct}
\delta_{gct} \phi(x) = \xi^{\m}(x)\pa_{\m} \phi(x).
\ee
This amounts just to the redefinition of the basis, i.e. we view $\xi^{\m}(x)$ and $\l^{\hA \hB}(x)$ as independent transformation parameters. Lorentz rotations act on fields with frame indices only, whereas general coordinate transformations are implemented by Lie derivative. 

General coordinate transformations have undesirable property that the variation themselves (like \eqref{gct}) do not transform covariantly under gauge transformations. To fix it we further define 'covariant general coordinate transformations' acting on the gauge field $B_{\m}$ by adding a field dependent gauge transformation
\be
\delta_{cgct}(\xi) = \delta_{gct}(\xi) - \delta(\xi^{\m} B_{\m}).
\ee

This might look a bit too abstract but at the end the result is very simple. The frame field transforms according to
\be
\delta_{cgct}(\x) e_{\m}^a = \pa_{\m}\xi^a + \xi^c B_{\m}^B f_{Bc}{^a} - \xi^{\n} T_{\m\n}^a,
\ee
where $T_{\m\n}^a$ is the torsion tensor and $f$ encodes structure constants of the gauged algebra. Whereas under the gauge transformations
\be
\delta e_{\m}^a = \pa_{\m}\xi^a + \xi^c B_{\m}^B f_{Bc}{}^a + \e^C(B_{\m}^B f_{BC}{}^a + e_{\m}^b f_{bC}{}^a).
\ee
Thus, if the torsion vanishes, then the cgct is the appropriate modification of local translation. 

Under diffeomorphisms and local Lorentz rotations the frame field and the spin connection transform according to
\al{
\delta E_{\m}^{\hA} &= \lcal_{\X} E_{\m}^{\hA} - \L^{\hA \hB}E_{\m \hB}, \\
\delta \O_{\m}^{\hA \hB} &= \lcal_{\X}\O_{\m}^{\hA \hB} +\pa_{\m}\L^{\hA \hB} + 2 \O_{\m}^{\hat{C} [\hA} \L^{\hB]}{}_{\hat{C}}.
}
Computing the commutator of two such transformation we obtain the algebra
\be
[(\X_2, \L_2),(\X_1,\L_1)] = ([\X_2, \X_1], [\L_2,\L_1] + \lcal_{\X_2}\L_1 - \lcal_{\X_1}\L_2).
\ee
For covariant general coordinate transformations the rules are:
\al{
\delta E_{\m}^{\hA} &= \pa_{\m} \X^{\hA} - \L^{\hA \hB}E_{\m \hB} + \X_{\hB} \O_{\m}^{\hA \hB} + \X^{\n} T_{\n \m}^{\hA}, \\
\delta \O_{\m}^{\hA \hB} &= \pa_{\m}\L^{\hA \hB} + 2 \O_{\m}^{\hat{C} [\hA} \L^{\hB]}{}_{\hat{C}}+ \X^{\n} R_{\n \m}^{\hA \hB}.
}
The corresponding algebra closes only if we impose zero torsion constraint and reads
\be
[(\X_2, \L_2),(\X_1,\L_1)] = (\L_2^{\hA\hB}\X_{1\hB}- \L_1^{\hA\hB}\X_{2\hB}, [\L_2,\L_1]^{\hA\hB}).
\ee


\bibliographystyle{JHEP}
\bibliography{literature}

\providecommand{\href}[2]{#2}\begingroup\raggedright\begin{thebibliography}{10}

\bibitem{Christensen:2013rfa}
M.~H. Christensen, J.~Hartong, N.~A. Obers and B.~Rollier, \emph{{Boundary
  Stress-Energy Tensor and Newton-Cartan Geometry in Lifshitz Holography}},
  \href{https://doi.org/10.1007/JHEP01(2014)057}{\emph{JHEP} {\bfseries 01}
  (2014) 057}, [\href{https://arxiv.org/abs/1311.6471}{{\ttfamily 1311.6471}}].

\bibitem{Hartong:2014oma}
J.~Hartong, E.~Kiritsis and N.~A. Obers, \emph{{Lifshitz space-times for
  Schrödinger holography}},
  \href{https://doi.org/10.1016/j.physletb.2015.05.010}{\emph{Phys. Lett.}
  {\bfseries B746} (2015) 318--324},
  [\href{https://arxiv.org/abs/1409.1519}{{\ttfamily 1409.1519}}].

\bibitem{Duval:2014uva}
C.~Duval, G.~W. Gibbons and P.~A. Horvathy, \emph{{Conformal Carroll groups and
  BMS symmetry}},
  \href{https://doi.org/10.1088/0264-9381/31/9/092001}{\emph{Class. Quant.
  Grav.} {\bfseries 31} (2014) 092001},
  [\href{https://arxiv.org/abs/1402.5894}{{\ttfamily 1402.5894}}].

\bibitem{Duval:2014lpa}
C.~Duval, G.~W. Gibbons and P.~A. Horvathy, \emph{{Conformal Carroll groups}},
  \href{https://doi.org/10.1088/1751-8113/47/33/335204}{\emph{J. Phys.}
  {\bfseries A47} (2014) 335204},
  [\href{https://arxiv.org/abs/1403.4213}{{\ttfamily 1403.4213}}].

\bibitem{Banados:2006fe}
M.~Banados, O.~Miskovic and S.~Theisen, \emph{{Holographic currents in first
  order gravity and finite Fefferman-Graham expansions}},
  \href{https://doi.org/10.1088/1126-6708/2006/06/025}{\emph{JHEP} {\bfseries
  06} (2006) 025}, [\href{https://arxiv.org/abs/hep-th/0604148}{{\ttfamily
  hep-th/0604148}}].

\bibitem{Klemm:2007yu}
D.~Klemm and G.~Tagliabue, \emph{{The CFT dual of AdS gravity with torsion}},
  \href{https://doi.org/10.1088/0264-9381/25/3/035011}{\emph{Class. Quant.
  Grav.} {\bfseries 25} (2008) 035011},
  [\href{https://arxiv.org/abs/0705.3320}{{\ttfamily 0705.3320}}].

\bibitem{Petkou:2010ve}
A.~C. Petkou, \emph{{Torsional degrees of freedom in AdS4/CFT3}},  2010,
  \href{https://arxiv.org/abs/1004.1640}{{\ttfamily 1004.1640}},
  \href{https://inspirehep.net/record/851578/files/arXiv:1004.1640.pdf}{https://inspirehep.net/record/851578/files/arXiv:1004.1640.pdf}.

\bibitem{Blagojevic:2013bu}
M.~Blagojevic, B.~Cvetkovic, O.~Miskovic and R.~Olea, \emph{{Holography in 3D
  AdS gravity with torsion}},
  \href{https://doi.org/10.1007/JHEP05(2013)103}{\emph{JHEP} {\bfseries 05}
  (2013) 103}, [\href{https://arxiv.org/abs/1301.1237}{{\ttfamily 1301.1237}}].

\bibitem{Cvetkovic:2017fxa}
B.~Cvetkovic, O.~Miskovic and D.~Simic, \emph{{Holography in Lovelock
  Chern-Simons AdS Gravity}},
  \href{https://doi.org/10.1103/PhysRevD.96.044027}{\emph{Phys. Rev.}
  {\bfseries D96} (2017) 044027},
  [\href{https://arxiv.org/abs/1705.04522}{{\ttfamily 1705.04522}}].

\bibitem{deHaro:2000vlm}
S.~de~Haro, S.~N. Solodukhin and K.~Skenderis, \emph{{Holographic
  reconstruction of space-time and renormalization in the AdS / CFT
  correspondence}}, \href{https://doi.org/10.1007/s002200100381}{\emph{Commun.
  Math. Phys.} {\bfseries 217} (2001) 595--622},
  [\href{https://arxiv.org/abs/hep-th/0002230}{{\ttfamily hep-th/0002230}}].

\bibitem{Fefferman:2007rka}
C.~Fefferman and C.~R. Graham, \emph{{The ambient metric}},
  \href{https://arxiv.org/abs/0710.0919}{{\ttfamily 0710.0919}}.

\bibitem{Hartong:2015usd}
J.~Hartong, \emph{{Holographic Reconstruction of 3D Flat Space-Time}},
  \href{https://doi.org/10.1007/JHEP10(2016)104}{\emph{JHEP} {\bfseries 10}
  (2016) 104}, [\href{https://arxiv.org/abs/1511.01387}{{\ttfamily
  1511.01387}}].

\bibitem{Grumiller:2016pqb}
D.~Grumiller and M.~Riegler, \emph{{Most general AdS$_{3}$ boundary
  conditions}}, \href{https://doi.org/10.1007/JHEP10(2016)023}{\emph{JHEP}
  {\bfseries 10} (2016) 023},
  [\href{https://arxiv.org/abs/1608.01308}{{\ttfamily 1608.01308}}].

\bibitem{Brown:1986nw}
J.~D. Brown and M.~Henneaux, \emph{{Central Charges in the Canonical
  Realization of Asymptotic Symmetries: An Example from Three-Dimensional
  Gravity}}, \href{https://doi.org/10.1007/BF01211590}{\emph{Commun. Math.
  Phys.} {\bfseries 104} (1986) 207--226}.

\bibitem{Skenderis:1999nb}
K.~Skenderis and S.~N. Solodukhin, \emph{{Quantum effective action from the AdS
  / CFT correspondence}},
  \href{https://doi.org/10.1016/S0370-2693(99)01467-7}{\emph{Phys. Lett.}
  {\bfseries B472} (2000) 316--322},
  [\href{https://arxiv.org/abs/hep-th/9910023}{{\ttfamily hep-th/9910023}}].

\bibitem{Imbimbo:1999bj}
C.~Imbimbo, A.~Schwimmer, S.~Theisen and S.~Yankielowicz,
  \emph{{Diffeomorphisms and holographic anomalies}},
  \href{https://doi.org/10.1088/0264-9381/17/5/322}{\emph{Class. Quant. Grav.}
  {\bfseries 17} (2000) 1129--1138},
  [\href{https://arxiv.org/abs/hep-th/9910267}{{\ttfamily hep-th/9910267}}].

\bibitem{Aksteiner:2015uxw}
S.~Aksteiner and Y.~Korovin, \emph{{New modes from higher curvature corrections
  in holography}}, \href{https://doi.org/10.1007/JHEP03(2016)166}{\emph{JHEP}
  {\bfseries 03} (2016) 166},
  [\href{https://arxiv.org/abs/1511.08747}{{\ttfamily 1511.08747}}].

\bibitem{Nakayama:2013is}
Y.~Nakayama, \emph{{Scale invariance vs conformal invariance}},
  \href{https://doi.org/10.1016/j.physrep.2014.12.003}{\emph{Phys. Rept.}
  {\bfseries 569} (2015) 1--93},
  [\href{https://arxiv.org/abs/1302.0884}{{\ttfamily 1302.0884}}].

\bibitem{Nakayama:2012sn}
Y.~Nakayama, \emph{{Holographic Renormalization of Foliation Preserving Gravity
  and Trace Anomaly}},
  \href{https://doi.org/10.1007/s10714-012-1427-3}{\emph{Gen. Rel. Grav.}
  {\bfseries 44} (2012) 2873--2889},
  [\href{https://arxiv.org/abs/1203.1068}{{\ttfamily 1203.1068}}].

\bibitem{Nakayama:2009qu}
Y.~Nakayama, \emph{{Forbidden Landscape from Holography}},
  \href{https://doi.org/10.1088/1126-6708/2009/11/061}{\emph{JHEP} {\bfseries
  11} (2009) 061}, [\href{https://arxiv.org/abs/0907.0227}{{\ttfamily
  0907.0227}}].

\bibitem{deWit:1981vgr}
B.~de~Wit, \emph{{CONFORMAL INVARIANCE IN EXTENDED SUPERGRAVITY}},  in
  \emph{{First School on Supergravity Trieste, Italy, April 22-May 6, 1981}},
  p.~0267, 1981.

\bibitem{Didenko:2012vh}
V.~E. Didenko and E.~D. Skvortsov, \emph{{Towards higher-spin holography in
  ambient space of any dimension}},
  \href{https://doi.org/10.1088/1751-8113/46/21/214010}{\emph{J. Phys.}
  {\bfseries A46} (2013) 214010},
  [\href{https://arxiv.org/abs/1207.6786}{{\ttfamily 1207.6786}}].

\bibitem{Kachru:2008yh}
S.~Kachru, X.~Liu and M.~Mulligan, \emph{{Gravity duals of Lifshitz-like fixed
  points}}, \href{https://doi.org/10.1103/PhysRevD.78.106005}{\emph{Phys. Rev.}
  {\bfseries D78} (2008) 106005},
  [\href{https://arxiv.org/abs/0808.1725}{{\ttfamily 0808.1725}}].

\bibitem{Taylor:2008tg}
M.~Taylor, \emph{{Non-relativistic holography}},
  \href{https://arxiv.org/abs/0812.0530}{{\ttfamily 0812.0530}}.

\bibitem{AyonBeato:2009nh}
E.~Ayon-Beato, A.~Garbarz, G.~Giribet and M.~Hassaine, \emph{{Lifshitz Black
  Hole in Three Dimensions}},
  \href{https://doi.org/10.1103/PhysRevD.80.104029}{\emph{Phys. Rev.}
  {\bfseries D80} (2009) 104029},
  [\href{https://arxiv.org/abs/0909.1347}{{\ttfamily 0909.1347}}].

\bibitem{Cai:2009ac}
R.-G. Cai, Y.~Liu and Y.-W. Sun, \emph{{A Lifshitz Black Hole in Four
  Dimensional R**2 Gravity}},
  \href{https://doi.org/10.1088/1126-6708/2009/10/080}{\emph{JHEP} {\bfseries
  10} (2009) 080}, [\href{https://arxiv.org/abs/0909.2807}{{\ttfamily
  0909.2807}}].

\bibitem{Taylor:2015glc}
M.~Taylor, \emph{{Lifshitz holography}}, {\emph{Class. Quant. Grav.} {\bfseries
  33} (2016) 033001}, [\href{https://arxiv.org/abs/1512.03554}{{\ttfamily
  1512.03554}}].

\bibitem{Ross:2011gu}
S.~F. Ross, \emph{{Holography for asymptotically locally Lifshitz spacetimes}},
  \href{https://doi.org/10.1088/0264-9381/28/21/215019}{\emph{Class. Quant.
  Grav.} {\bfseries 28} (2011) 215019},
  [\href{https://arxiv.org/abs/1107.4451}{{\ttfamily 1107.4451}}].

\bibitem{Mann:2006bd}
R.~B. Mann, D.~Marolf and A.~Virmani, \emph{{Covariant Counterterms and
  Conserved Charges in Asymptotically Flat Spacetimes}},
  \href{https://doi.org/10.1088/0264-9381/23/22/017}{\emph{Class. Quant. Grav.}
  {\bfseries 23} (2006) 6357--6378},
  [\href{https://arxiv.org/abs/gr-qc/0607041}{{\ttfamily gr-qc/0607041}}].

\bibitem{Mann:2008ay}
R.~B. Mann, D.~Marolf, R.~McNees and A.~Virmani, \emph{{On the Stress Tensor
  for Asymptotically Flat Gravity}},
  \href{https://doi.org/10.1088/0264-9381/25/22/225019}{\emph{Class. Quant.
  Grav.} {\bfseries 25} (2008) 225019},
  [\href{https://arxiv.org/abs/0804.2079}{{\ttfamily 0804.2079}}].

\bibitem{Compere:2011db}
G.~Compere, F.~Dehouck and A.~Virmani, \emph{{On Asymptotic Flatness and
  Lorentz Charges}},
  \href{https://doi.org/10.1088/0264-9381/28/14/145007}{\emph{Class. Quant.
  Grav.} {\bfseries 28} (2011) 145007},
  [\href{https://arxiv.org/abs/1103.4078}{{\ttfamily 1103.4078}}].

\bibitem{Freedman:2012zz}
D.~Z. Freedman and A.~Van~Proeyen, \emph{{Supergravity}}.
\newblock Cambridge Univ. Press, Cambridge, UK, 2012.

\end{thebibliography}\endgroup

\end{document}